\documentclass[aps,prd,eqsecnum,11pt,showpacs,nofootinbib]{revtex4-2}

\usepackage{dcolumn}
\usepackage{bm}
\usepackage{xcolor}
\usepackage[applemac]{inputenc}
\usepackage[spanish,english]{babel}
\usepackage{amsmath,amssymb,amsfonts,latexsym,cancel,amsthm,hyperref,bbm}
\usepackage{graphicx}
\usepackage{color}
\usepackage{soul}
\usepackage{ulem}
\usepackage{slashed}
\usepackage{braket}
\usepackage[mathscr]{euscript}
\usepackage{physics}
\usepackage{mathrsfs}
\usepackage{pgfplots}
\usepackage{float}
\usepackage{xfrac}
\usepackage{xcolor}
\usepackage{trimclip}
\usepackage{amsmath}
\usepackage{accents}
\definecolor{blue(pigment)}{rgb}{0.2, 0.2, 0.6}
\allowdisplaybreaks

\newcommand{\la}{\langle}
\newcommand{\ra}{\rangle}
\newcommand{\w}{\omega}

\newcommand{\be}{\begin{equation}}
\newcommand{\ee}{\end{equation}}
\newcommand{\bea}{\begin{eqnarray}}
\newcommand{\eea}{\end{eqnarray}}
\newcommand{\bes}
{\begin{subequations}}
\newcommand{\ees}{\end{subequations}}

\begin{document}
 \title{\bf Vacuum polarization and stress-energy of a quantum field inside of two-dimensional black holes }

   \author{Paul R. Anderson}\email{anderson@wfu.edu}
    \affiliation{Department of Physics, Wake Forest University, Winston-Salem, NC, 27109, USA}

   \author{Amanda Peake}
    \affiliation{Department of Physics, Wake Forest University, Winston-Salem, NC, 27109, USA}

 \author{Shohreh Gholizadeh Siahmazgi}\email{gholizs@wfu.edu}
    \affiliation{Department of Mathematics, Wake Forest University, Winston-Salem, NC, 27109, USA}
    \affiliation{Department of Physics, Wake Forest University, Winston-Salem, NC, 27109, USA}
    
\begin{abstract}
Quantum effects are studied in both Schwarzschild spacetime and a spacetime in which a null shell collapses to form a black hole via the vacuum polarization $\la \phi^2 \ra$ and stress-energy tensor $\la T_{ab} \ra$ for a massless minimally-coupled scalar field in two dimensions.  For Schwarzschild spacetime, the Boulware, Unruh, and Hartle-Hawking states are considered. For the collapsing null shell spacetime, the {\it in} vacuum state is used.  Instabilities of the Unruh, Hartle-Hawking, and {\it in} states resulting from the behavior of $\la \phi^2 \ra$ in the regions inside and outside of the horizon are found.  The question of how well the Unruh state for the eternal black hole approximates quantum effects in the interior of a black hole that forms from collapse is addressed.

   \end{abstract}

    \date{\today}
    \maketitle

\newpage
\section{Introduction}

The three most important states for eternal black holes are the Boulware state~\cite{Boulware}, the Hartle-Hawking state~\cite{H-H}, and the Unruh state~\cite{Unruh}. The Boulware state is the true vacuum state at both past and future null infinity, but the stress-energy of a quantum field diverges on the past and future horizons in this state.  The Hartle-Hawking state best describes a black hole in thermal equilibrium with its Hawking radiation in a cavity surrounded by a perfectly reflecting mirror.  The stress-energy is finite on the past and future horizons.  It can be used to study black hole thermodynamics.  The Unruh state is the best state to approximate Hawking radiation for an eternal black hole.  The stress-energy is finite on the future horizon and there is a flux of radiation in a thermal distribution at the Hawking temperature going out to future null infinity. The stress-energy does diverge on the past horizon for the Unruh state but this is not important because there is no past horizon for a black hole that forms from collapse. 
For a black hole that forms from collapse in a spherically symmetric, asymptotically flat spacetime, the {\it in} state is defined to be the vacuum state at past null infinity along with the condition that before the black hole forms the mode functions are regular at $r = 0$.  

While these states have been studied extensively in both two-dimensions, 2D, and four-dimensions, 4D, most of what is known about them for Schwarzschild black holes is for the region outside the past and future horizons which we call the exterior region and for the horizons that bound that region.  
In this paper, we work with a massless minimally-coupled scalar field in 2D where analytic calculations are possible and investigate the behaviors of the vacuum polarization $\la \phi^2 \ra$ and the stress-energy tensor $\la T_{ab} \ra$ for all four states in Schwarzschild spacetime with a focus on the region inside the future horizon and outside the past horizon which we call the interior region.  

The vacuum polarization does not appear to have been studied previously in any 2D black hole spacetime. Previous calculations of the stress-energy tensor for a massless minimally-coupled scalar field in 2D Schwarzschild spacetime for all four states have been done and the results have been analyzed in the exterior region~\cite{Davies-Fulling-Unruh, Fulling, Hiscock, Sandro-Book}, but not in the interior region with the exception of correlations between the energy density in the interior and exterior regions~\cite{Sandro-Roberto}.  There have been several full numerical calculations of the vacuum polarization~\cite{Fawcett-Whiting, Candelas-Howard, And-phi2,Levi-Ori-phi2-1,Levi-Ori-phi2-2} and the stress-energy tensor~\cite{Fawcett, Howard-Candelas, Howard, Jensen-McLaughlin-Ottewill-1, Jensen-McLaughlin-Ottewill-2,AHS-1, AHS-2, And-Balbinot-Fabbri,Levi-Ori-Tab, Taylor-Breen-Ottewill} for scalar fields in Schwarzschild spacetime in the Boulware, Unruh, and Hartle-Hawking states in the exterior region.  There have been two full numerical calculations of the vacuum polarization in part of the interior region of Schwarzschild spacetime in 4D in the Hartle-Hawking and Unruh states~\cite{Candelas-Jensen,Lanir-Levi-Ori-phi2-inside}.  Recently there has been a calculation of the vacuum polarization and one component of the stress-energy tensor near the trajectory of a massless spherically symmetric shell in 4D that collapses to form a black hole~\cite{ Ori-Zilberman}. 

One reason that it is interesting to investigate the behavior of the vacuum polarization is that it was previously found that, for a massless minimally-coupled scalar field in the Unruh state, the symmetric two-point function or Hadamard Green's function grows without bound in time at late times in the static region outside of an eternal 2D black hole when the points are split in the radial direction~\cite{A-T}.  The vacuum polarization can be used to test the robustness of the previous result since it is a function of just one spacetime point so there is no ambiguity associated with the way in which the points are split.   

The existence of the linear growth in time of the two-point function appears to be indicative of some type of instability associated with the Unruh state.  There is also evidence that it is associated with an infrared divergence in the two-point function which is caused by infrared divergences in the mode functions~\cite{A-G-S}. 
There is an interesting exception which is the collapsing null-shell spacetime in two dimensions.  For the model studied in~\cite{A-T} the mode functions consist of differences between two solutions to the mode equation that both have infrared divergences, however these divergences cancel. Nevertheless, at late times the same linear growth appears to leading order as was found for the Unruh state.  There is also a subleading term which grows logarithmically in time and is not present for the Unruh state.  
In 4D one would expect the linear growth to occur in spacetimes where scattering effects do not remove the infrared divergence in the mode functions.  Such growth has been found for a massless minimally-coupled scalar field in the Unruh state in 4D Schwarzschild-de Sitter spacetime in the static region between the black hole and cosmological horizons~\cite{N-P-A}.

In this paper, we find the same linear growth in time for the vacuum polarization that was found previously in~\cite{A-T} for the symmetric two-point function occurs in the exterior region of 2D Schwarzschild spactime and the region outside the shell and the horizon of a null shell that collapses to form a 2D black hole.  We also find similar behaviors in the interior region and a different type of divergent behavior at infinity for the Hartle-Hawking state in the exterior region.
 We also extend previous calculations of the stress-energy tensor for all four states into the region inside the horizon.  We compare both the vacuum polarization and the stress-energy in the null shell spacetime in both the interior and exterior regions with those in Schwarzschild spacetime when the field is in the Unruh state.  This has previously been done for the stress-energy in the exterior region but not otherwise.  The question of how good an approximation the Unruh state is to the {\it in} state is important because it is much easier in 4D to compute both the vacuum polarization and the stress-energy tensor for an eternal black hole than it is to compute them if the black hole forms from collapse.
 
In Sec. II, we give a brief overview of Schwarzschild spacetime and the relevant coordinates. In Sec. III, the complete sets of solutions to the mode equation that are used to define the Boulware, Hartle-Hawking, Unruh, and {\it in} states are given.  In Sec. IV, the quantity $\la \phi^2 \ra$ is computed and discussed for the Boulware, Hartle-Hawking, Unruh, and {\it in} states and its behavior is investigated for each.  A comparison is made between $\la \phi^2 \ra$ for the Unruh and {\it in} states.  In Sec. V the stress-energy tensor for the scalar field is analyzed for all four states in the interior region.  Again a comparison is made of its behaviors for the Unruh state and the {\it in} state.  Sec. VI contains a summary and discussion of our results.  A derivation of the stress-energy tensor in both the interior and exterior regions for all four states is given in the Appendix.  Throughout, we use units such that $\hbar = c = G = 1$.  

\section{Spacetimes Considered Here}
\subsection{Schwarzschild Spacetime
}
One of the spacetimes we consider is a 2D Schwarzschild black hole with the line element 
\begin{equation}\label{2D Schwarzchild Metric}
ds^2 = -f(r)dt^2 + \frac{dr^2}{f(r)}  \;,
\end{equation}
where  $f(r) = 1 - \frac{2M}{r} $
and $M$ represents the mass of the BH.  The tortoise coordinate is
\begin{equation}\label{r star}
    r_* = \int^r\frac{dr'}{f(r')} = r + 2M\log\bigg|1-\frac{r}{2M}\bigg|.
\end{equation}
In the interior region, $t$ is a space coordinate.  Both $r$ and $r_*$ can serve as time coordinates because they increase as one goes from the horizon to the spacelike singularity at $r=0$.

\begin{figure}[H]
    \centering
    \includegraphics[width=0.7\linewidth]{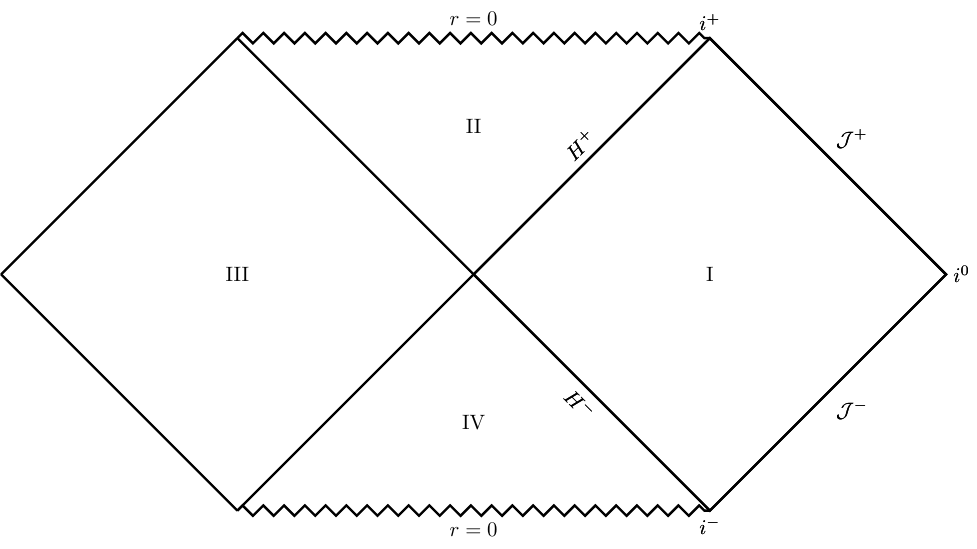}
    \caption{Penrose diagram for a Schwarzschild black hole.  The wavy lines are the singularities at $r = 0$.  The past and future horizons are labeled $H^\pm$ and past and future null infinity are labeled $\mathscr{I}^{\pm}$. We call the region labeled I, the exterior region and the region labeled II the interior region.}
    \label{fig:Schwarzschild}
\end{figure}

The Penrose diagram for Schwarzschild spacetime is given in Fig.\;\ref{fig:Schwarzschild}.

It is useful to define left and right moving null coordinates $u$ and $v$.  For the interior and exterior regions
\bes \bea \label{u int}
    u_{\text{int}} &=& r_* - t  \;, \\
\label{u}
    u_{\text{ext}} &=& t-r_* \;, \eea 
and in both regions
\be v = r_* + t \;. \ee \ees 
Note that the coordinate $v$ is regular on the future horizon while $u_{\rm ext} \to +\infty$ and $u_{\rm int} \to -\infty$ there.  The coordinates $u_{\rm int}$ and $u_{\rm ext}$ are regular on the past horizon in the interior and exterior regions respectively while $v \to -\infty$ everywhere on the past horizon.

The right-moving null Kruskal coordinate in the interior and exterior regions is
\bes \bea
    U(u_{\text{int}}) &=& 4 M \, e^{\frac{u_{\text{int}}}{4 M}} \; , \\
    U(u_{\text{\text{ext}}}) &=& -4 M\, e^{-\frac{u_{\text{ext}}}{4 M}} \;,
    \label{Kruskal-U}  \eea
and the left-moving Kruskal coordinate in both regions is
\be V(v) = 4 M \, e^{\frac{v}{4M}} \;. \ee \ees 
These coordinates are well-behaved throughout the spacetime.

\subsection{Collapsing Null Shell Spacetime}

We take the 2D analog of a 4D spherically symmetric massless shell that contracts along the null ray $v = v_0$ for any $-\infty < v_0 < \infty$.
The metric inside the shell is the flat space metric given by
\begin{equation}
    ds^2 = -dt_F^2 + dr^2. \label{flat-metric}
\end{equation}
In 2D one can have $-\infty < r < \infty$.  However, to approximate the 4D case where $r \ge 0$ one can put a perfectly reflecting mirror at $r = 0$ and just consider the spacetime outside the mirror.  That is what we do here.

In the flat space region inside the shell we define coordinates
\begin{equation}\label{flat space coordinate}
    u_F = t_F - r, \quad v = t_F + r,
\end{equation}
where $t_F$ is the flat space time coordinate. 
The metric outside the shell is given by the Schwarzchild metric \eqref{2D Schwarzchild Metric}. The two metrics are matched along the shell trajectory, which is a null trajectory along $v$= $v_0$. The matching is set up so that the coordinates $r$ and $v$ are continuous across the trajectory, but $t$ and $u$ are not. 
In the region outside the shell and the horizon, we can write $u_{\text{ext}}$ in terms of $u_F$ by  matching along the trajectory $v = v_0$. 
The result is~\cite{Sandro-Book} 

\begin{eqnarray} \label{uext-uf}
    u_{\rm ext} = u_F - 4 M \log \left( \frac{v_H - u_F}{4M}\right).
\end{eqnarray}
Inverting this gives~\cite{G-A-E}
\begin{equation} \label{uf-ext}
    u_F(u_{\text{ext}}) = v_H - 4MW\Big[\exp\Big(\frac{v_H-u_{\text{ext}}}{4M}\Big)\Big],
\end{equation}
where $v_H =v_0-4M$ and $W$ is the Lambert W function.
Following the same process as above, we can use 
$u_{\text{int}}$ as defined in Eq. (\ref{u int}) in the region outside the shell and inside of the horizon with the result

\begin{eqnarray} \label{uint-uF}
    u_{\rm int} = - u_F + 4 M \log \left( \frac{ u_F - v_H}{4M}\right)
\end{eqnarray}
and
\begin{equation}\label{uF-int}
     u_F(u_{\text{int}}) = v_H - 4MW\Big[-\exp\Big(\frac{u_{\text{int}}+v_H}{4M}\Big)\Big].  
\end{equation}.
\par
These coordinates have the following properties: 
\begin{enumerate}
    \item The event horizon of the BH within the flat space region is at $u_F = v_H =v_0 - 4M$.
    \item The null shell trajectory reaches the singularity $r=0$ at $u_F = v_0$. Therefore, the part of the trajectory in the interior region of the BH is $v_H < u_F \le v_0$. 
\end{enumerate}
\begin{figure}[H]
    \centering
  \includegraphics[trim=3cm 18cm 4cm 3cm,clip=true,totalheight=0.3\textheight]
    {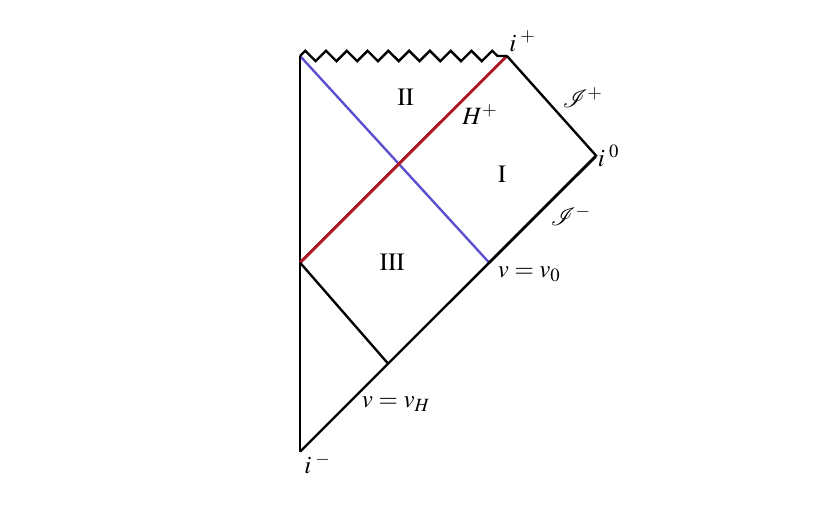}    \caption{Penrose Diagram of a black hole that forms from a collapsing null shell in 4D or in 2D when a perfectly reflecting mirror is placed at $r = 0$ and only the region to the right of the mirror is considered.  The vertical line corresponds to the spatial point $r = 0$ inside the shell and the wavy horizontal line corresponds to the singularity at $r=0$ outside the shell.  }
    \label{fig:Penrose Collapsing Shell}
\end{figure}
The Penrose diagram for a BH formed from a collapsing null shell is shown in Fig.\;\ref{fig:Penrose Collapsing Shell}.

\section{Mode Equation}

We consider a massless minimally-coupled scalar field $\phi$ satisfying 
\begin{equation}\label{box}
\Box \phi = \frac{1}{\sqrt{-g}}\partial_\alpha(g^{\alpha\beta}\sqrt{-g}\partial_\beta)\phi=0.
\end{equation}
For the flat space metric~\eqref{flat-metric}, one finds
\be \frac{\partial^2\phi}{\partial t^2}-\frac{\partial^2\phi}{\partial r^2}=0 \;.  \ee
The general solution is of form
\be \phi = g_F(u_F) + h_F(v) \;, \ee
where $g_F$ and $h_F$ are arbitrary functions.
For the Schwarzschild metric~\eqref{2D Schwarzchild Metric},  
\begin{equation}\label{box 2D}
    \frac{\partial^2\phi}{\partial t^2}-\frac{\partial^2\phi}{\partial r_*^2}=0
\end{equation}
which has a general solution of the form
\begin{equation} \label{f and g}
    \phi = g(u_i) + h(v),
\end{equation}
where $g$ and $h$ are arbitrary functions.  In the exterior region
\be u_i = u_{\rm ext} \;, \ee
and in the interior region 
\be u_i = u_{\rm int} \;. \ee  

A complete set of orthonormal mode functions that are solutions to Eq. (\ref{box}) defines a `vacuum state' of the field $\phi$. For the states we consider, each complete set can be broken up into separate sets of mode functions that we denote by $f^i_\w$ with $\w \ge 0$.  Then
\be \phi = \int_{0}^\infty d \w \sum_i [a^i_\w f^i_\w + a^{i \, \dagger}_\w f^{i \, *}_\w ] \;. \label{phi-gen} \ee  
The modes are normalized using the usual scalar product~\cite{Sandro-Book}. 

\subsection{Important vacuum states in Schwarzschild spacetime}

The Boulware, Unruh, and Hartle-Hawking states are composed of various combinations of two complete sets of modes:  the Boulware modes and the Kruskal modes.  We begin by discussing these two sets of modes.  Our discussion will only focus on the behaviors of these modes in interior and exterior regions.  Because there is no scattering for the massless minimally-coupled scalar field in two dimensions, we can write down analytic expressions for these modes everywhere in these regions.

The Boulware modes define what is a true vacuum state at past and future null infinity.  They are broken up into three subsets, those that have initial data given on the past horizon in the exterior region, which we denote by $h^{b \, {\rm ext}}_\w$, those with initial data on the past horizon in the interior region, which are denoted by $h^{b \, {\rm int}}_\w$, and those with initial data on past null infinity which are denoted by $h^{\mathscr{I}^{-}}_\w$.  In the exterior region
\bea  h^{b \, {\rm ext}}_\w &=& \frac{1}{\sqrt{4 \pi \w}} e^{-i \w u_{\rm ext}},
\nonumber \\
      h^{b \, {\rm int}}_\w &=&  0,  \nonumber \\
    h^{\mathscr{I}^{-}}_\w &=& \frac{1}{\sqrt{4 \pi \w}} e^{-i \w v}  \;.  \eea
In the interior region,
\bea  h^{b \, {\rm ext}}_\w &=& 0, \nonumber \\
      h^{b \, {\rm int}}_\w &=& \frac{1}{\sqrt{4 \pi \w}} e^{i \w u_{\rm int}}, \nonumber \\
      h^{\mathscr{I}^{-}}_\w &=& \frac{1}{\sqrt{4 \pi \w}} e^{-i \w v}  \;.  \eea
    
The Kruskal modes are used in the definitions of both the Unruh and Hartle-Hawking states. Those denoted by $p^{H^-}_\w$ have initial data on the past horizon $H^-$ and those denoted by  $p^{H^+}_\w$ have initial data on the future horizon $H^+$.  In the interior and exterior regions they have the forms
\bea p^{H^-}_\w &=& \frac{1}{\sqrt{4 \pi \w}} e^{-i \w U(u_i)},  \nonumber \\
     p^{H^+}_\w &=& \frac{1}{\sqrt{4 \pi \w}} e^{-i \w V(v)} \;. \eea

\subsection{Vacuum state for the null shell spacetime}

The natural {\it in} vacuum state for the massless minimally-coupled scalar field in the null shell spacetime can be specified by requiring that on past null infinity
\be f^{\rm in}_\w = \frac{e^{-i \w v}}{\sqrt{4 \pi \w}} \ee
and inside the shell at $r = 0$, the mode functions must vanish due to the existence of the perfectly reflecting mirror.  These two conditions result in the following set of mode functions for the {\it in} state both inside and outside the null shell
\be f^{\rm in}_\w = \frac{1}{\sqrt{4 \pi \w}} (e^{-i \w v} - e^{-i \w u_F} ) \;. \label{fin} \ee
Outside the null shell and the horizon $u_F = u_F(u_{\text{ext}})$ is given by the relation~\eqref{uf-ext}.  Outside the null shell and inside the horizon, $u_F = u_F(u_{\rm int})$ is given by the relation~\eqref{uF-int}. 

\section{Two-point function and $\la \phi^2 \ra$ }

\subsection{Schwarzschild spacetime}

The symmetric two-point correlation function for a scalar field is given by
\be G^{(1)}(x,x') \equiv \la 0| [\phi(x) \phi(x') + \phi(x') \phi(x)] |0 \ra \;. \label{G1-def} \ee
Substituting~\eqref{phi-gen} into this equation gives
\be G^{(1)}(x,x') = \int_0^\infty d \w \sum_i [f^i_\w(x) f^{i \, *}_\w(x') + f^i_\w(x') f^{i \, *}_\w(x)] \;. \label{G1-gen} \ee

For each of the sets of mode functions comprising the Boulware, Unruh, and Hartle-Hawking states, there is an infrared divergence of the form $\frac{1}{\w}$ in the integrand, so we replace the lower limit with an infrared cutoff $\lambda$. For the Boulware state, in Region I outside the horizon,
\bea \label{Green's Boulware ext} G_{B \,\rm ext}^{(1)}(x,x') &=& 
      \int_\lambda^\infty d\omega\bigg[\frac{1}{4\pi\omega}\Big( e^{-i\omega (u_{\text{ext}}-u_{\text{ext}}')} + e^{i\omega (u_{\text{ext}}-u_{\text{ext}}')} 
     \\ \nonumber  &&
     + e^{-i\omega (v-v')} + e^{i\omega(v-v')}\Big)\bigg] \nonumber \\
     &=&  -\frac{1}{2 \pi} \{ {\rm ci}[\lambda (u_{\rm ext} - u'_{\rm ext})] - {\rm ci}[\lambda (v-v')]\} \;.
\end{eqnarray} 
In the interior region,
\begin{eqnarray}\label{Boulware-2-point}
 G^{(1)}_{B \, {\rm int}}(x,x') &=& \int_\lambda^\infty d\omega\bigg[\frac{1}{4\pi\omega}\Big( e^{i\omega (u_{\text{int}}-u_{\text{int}}')} + e^{-i\omega (u_{\text{int}}-u_{\text{int}}')} 
     \\ \nonumber  &&
     + e^{i\omega (v-v')} + e^{-i\omega(v-v')}\Big)\bigg] \nonumber \\
     &=&  -\frac{1}{2 \pi} \{{\rm ci}[\lambda (u_{\text{int}}-u'_{\text{int}})] + {\rm ci}[\lambda (v-v')]\} \;. 
\end{eqnarray}
Here ci is the usual cosine integral function.
One similarly finds for both regions that the expressions for the Unruh state and Hartle-Hawking states are
\begin{eqnarray}
 G^{(1)}_{U}(x,x') &=&  -\frac{1}{2 \pi} \{{\rm ci}[\lambda (U(u_i)-U(u'_i))] + {\rm ci}[\lambda (v-v')]\} \;, \label{G1-U} \\ 
 G^{(1)}_{H}(x,x') &=&  -\frac{1}{2 \pi} \{{\rm ci}[\lambda (U(u_i)-U(u'_i))] + {\rm ci}[\lambda [V(v)-V(v')]\} \;. \label{G1-HH}
\end{eqnarray}
Here $u_i = u_{\rm ext}$ in the exterior region and $u_i = u_{\rm int}$ in the interior region.

The formal relation between the vacuum polarization and the symmetric two-point function is
$\la \phi^2(x) \ra = \frac{1}{2} \lim_{x' \to x} G^{(1)}(x,x')$.  
However, there is a divergence when the two points come together, and thus this quantity has to be renormalized.  Here, we use the method of point splitting.  The point splitting counter term in the 2D case is given in~\cite{Bunch-Christensen-Fulling}.  In general, renormalization counter terms do not depend on the state but do depend on how the points are split.  Here, we assume that the points are split in the $t$ direction in both the interior and the exterior regions.  In the Appendix we show that the point splitting counter terms in this case are
\be \la \phi^2 \ra_{ps} =  -\frac{1}{2 \pi} \left\{ \gamma_E + \frac{1}{2} \log\left[ \frac{ |r- 2 M| \mu^2 (t-t')^2}{4 r} \right] \right\} \;, \label{phi2-ps} \ee
where $\gamma_{E}$ is Euler's constant and $\mu$ is an arbitrary constant with units of mass.
Then the renormalized expression is given by
\be \la \phi^2 \ra = \lim_{t' \to t} \left[ \frac{1}{2} G^{(1)}(t,r;t',r) - \la \phi^2 \ra_{ps} \right] \;. \label{phi2-ren} \ee

For the Boulware state, we use the expansion 
\be {\rm ci}(z) = \gamma_E + \log z + O(z^2) \label{cosintexp}\ee
to obtain
\be \langle\phi(x)^2\rangle_{B} = \frac{1}{4\pi}\log\Big(\frac{|r-2M|}{4r}\frac{\mu^2}{\lambda^2}\Big)
\label{phi2-B}
\ee
for both the interior and exterior regions.

For differences of the Kruskal modes, the expansion in powers of $t - t'$ when $r' =r$ is more subtle. Since the renormalization counter terms are expanded in terms of $t-t'$, it is necessary to have the same expansions for the unrenormalized expressions.  
Using Taylor series expansions, one finds 
\bes \bea
    U(u'_{\text{ext}}) &\approx& U(u_{\text{ext}}) + \left(\frac{dU}{du_{\text{ext}}} \right) (u'_{\text{ext}}-u_{\text{ext}}) =  U(u_{\text{ext}}) - e^{-\frac{u_{\text{ext}}}{4M}} (t-t')\;, \\
    U(u'_{\text{int}}) &\approx& U(u_{\text{int}}) + \left(\frac{dU}{du_{\text{int}}}\right) (u'_{\text{int}}-u_{\text{int}})  =     U(u_{\text{int}}) +  e^{ \frac{u_{\text{int}}}{4M}} (t-t')\;, \\
    V(v') &\approx& V(v) + \left(\frac{d V}{dv} \right) (v'-v) = V(v) - e^{\frac{ v}{4 M}} (t-t') \;. \eea \label{U-V-expansion}  \ees
    
Using these results in~\eqref{G1-U} and~\eqref{G1-HH} and combining with~\eqref{phi2-ps}, it is straightforward to show that 
\bea    
\langle\phi(x)^2\rangle_{H} &=& \frac{1}{4\pi}\left[- \frac{r_*}{2M} + \log\left(\frac{|r-2M|}{4r} \frac{\mu^2}{\lambda^2}\right)\right] = \frac{- 1}{8\pi M} r_* + \langle\phi(x)^2\rangle_{B} \nonumber \\ &=& - \frac{r}{8 \pi M} + \frac{1}{4 \pi} \log \left( \frac{M}{2 r} \frac{\mu^2}{\lambda^2} \right) \;,  \label{phi2-H}   \\
\langle\phi(x)^2\rangle_U &=& \frac{1}{4\pi}\left[\frac{t-r_*}{4 M} + \log \left(\frac{|r-2M|}{4r} \frac{\mu^2}{\lambda^2} \right) \right] = \frac {v}{16 \pi M} + \la \phi^2 \ra_H \nonumber \\
&=& \frac{u_{ext}}{16 \pi M} + \la \phi^2 \ra_B \;, \label{phi2-U}
\eea
where, except for the last equality for $\la \phi^2 \ra_U$, each expression is valid in both the interior and the exterior regions. 

From~\eqref{phi2-B}, it is clear that $\la \phi^2 \ra_B$ is independent of $t$ and vanishes in the limit $r \to \infty$ if $\mu^2 = 4 \lambda^2$. As expected for the Boulware state, it diverges on the past and future horizons. 

From~\eqref{phi2-H}, it can be seen that $\la \phi^2 \ra_H$ is also independent of $t$ and is finite on the past and future horizons, as expected for the Hartle-Hawking state.  What is surprising is that it has both a linear and logarithmic divergence in the limit $r \to \infty$.  

The behavior of $\la \phi^2 \ra_U$ is more complicated due to its dependence on $t$.  As can be seen from~\eqref{phi2-U}, it is finite on future null infinity for finite values of $u_{\text{ext}}$ as expected, but it diverges elsewhere in the limit $r \to \infty$.  On the future horizon, it is finite for all finite values of $v$ since $v$ is continuous across that horizon.  It diverges on the past horizon, as expected, since $v \to - \infty$ there.  However, it also diverges on the future horizon in the limit $v \to \infty$ and diverges at past and future timelike infinity, since it is proportional to $t$ for fixed values of $r$.  

In the interior region, $\la \phi^2 \ra$ diverges for all three states in the same way as $r \to 0$.  For the Boulware and Hartle-Hawking states, it monotonically decreases throughout the interior region. Because of the dependence on $t$, there is an inhomogeneity in $\la \phi^2 \ra_U$ in the interior region and a divergence for fixed $r$ in the limits $t \to \pm \infty$.

\begin{figure}[H]
    \centering
    \includegraphics[width=0.7\linewidth]{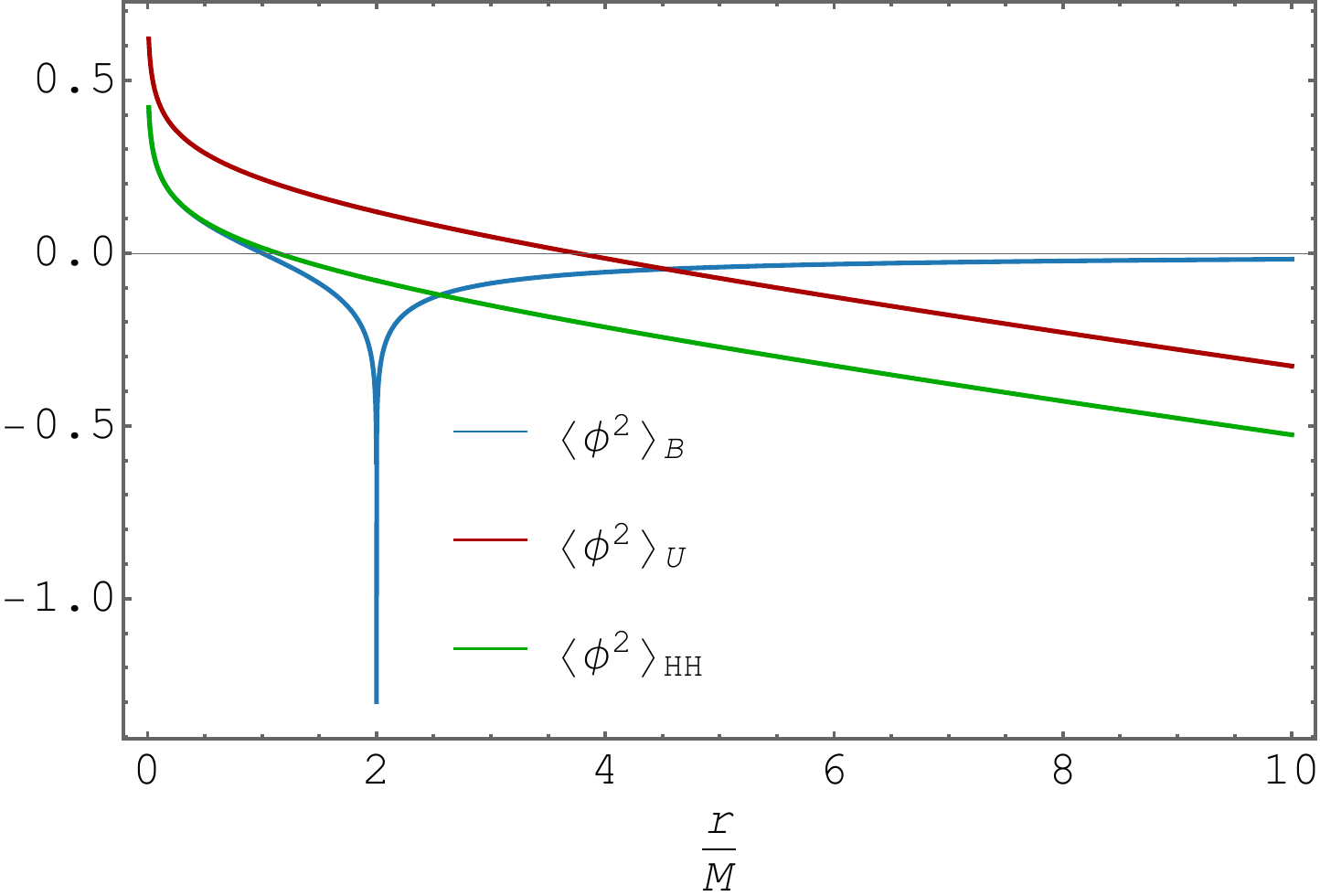}
    \caption{Plot of $\la \phi^2\ra$ as a function of $\frac{r}{M}$. On the right-hand side of the plot, from top to bottom, the curves are for the Boulware (blue), Unruh (red), and Hartle-Hawking (green) states.  For the Unruh state, the plot is for the surface $\frac{v}{M} = 10$.  The plot covers both the interior and the exterior regions.}
    \label{fig-3-states}
\end{figure}

Some of these behaviors are illustrated in Fig.~\ref{fig-3-states}.

\subsection{Null shell spacetime }

For the {\it in} vacuum state for the collapsing null shell spacetime there is no infrared divergence.  In this case it turns out to be easiest to compute $G^{(1)}$ using the cutoff and take the limit $\lambda \to 0$ at the end of the calculation.  Using the relation ${\rm ci}(z) = \gamma_E + \log z + O(z^2)$, we find in the parts of the interior and exterior regions that are outside the null shell trajectory, $v > v_0$,
\begin{eqnarray}
    G^{(1)}_{\rm in}(x,x') 
    &=& \frac{1}{2\pi} \lim_{\lambda \to 0} \int_\lambda^\infty \frac{d\omega}{\omega}  \left\{ 
    \cos[\omega (v-v')] + \cos[\omega (u_F(u_i)-u_F(u'_i))] \right. \nonumber \\
    &&  \left. \qquad - \cos[\omega (u_F(u_i)-v')] - \cos[\omega(v-u_F(u'_i)] \right\} \nonumber \\
    &=& \frac{1}{2\pi} \log \left\vert\frac{(u_F(u_i)-v')(v - u_F(u'_i))}{(v-v')(u_F(u_i)-u_F(u'_i))}\right\vert \;,
    \label{G1-null-shell}
\end{eqnarray}
where $u_i$ is defined above.

To renormalize $\la \phi^2 \ra_{in}$, we expand the unrenormalized expression in powers of $t-t'$ using Taylor series expansions which require the computation of $\frac{du_F}{d u_{\text{ext}}}$ and $\frac{du_F}{d u_{\text{int}}}$. 
These can be obtained from (\ref{uext-uf}) and (\ref{uint-uF}) with the result
\begin{eqnarray}
    \frac{du_F}{du_{\text{ext}}} &=& \frac{1}{\frac{d u _{\rm ext} }{d u_F}}
        = \Bigg(\frac{v_H-u_F}{v_0-u_F}\Bigg), \label{duf/(duext} \\
        \frac{du_F}{du_{\text{int}}} &=& \frac{1}{\frac{d u _{\rm int} }{d u_F}}
        = \Bigg(\frac{v_H-u_F}{u_F-v_0}\Bigg). \label{duf/(duint}
\end{eqnarray}
After some calculation, we find that 
\begin{eqnarray}
    \langle\phi(x)^2\rangle_{in} &=& \frac{1}{4\pi}\Bigg(2\gamma_E +
     \log\Bigg|\frac{(v_0-u_F(u_i)\big)}{\big(v_H-u_F(u_i)\big)}\Bigg|
    +  \log\Bigg|\big(u_F(u_i)-v\big)^2  \mu^2 \frac{(r-2M)}{4r} \Bigg|\Bigg) \nonumber \\ 
    &=&  \frac{1}{4\pi}\Bigg( 2\gamma_E +
     \log\Bigg|\frac{(v_0-u_F(u_i)\big)}{\big(v_H-u_F(u_i)\big)}\Bigg|
    +  \log \left[ \lambda^2 \big(u_F(u_i)-v\big)^2 \right] \Bigg) + \langle\phi(x)^2\rangle_{B}  \;, \label{phi2-ns}
\end{eqnarray}
where again $u_i = u_{\text{ext}}$ in the exterior region and $u_i = u_{\text{int}}$ in the interior region.
Note that there is no infrared divergence in the integrand for $G^{(1)}(x,x')$ in this case so no lower limit cutoff needs to be introduced. In the second line we have inserted a factor of this cutoff in order to show explicitly the relationship with $\la \phi^2 \ra_B$.  The full expression is independent of $\lambda$.

On future null infinity, $u_{\text{ext}}$ is a regular coordinate and thus $u_F(u_{\text{ext}})$ is as well.  However, $v = + \infty$ on this surface so the 
$\log |u_F(u_i)-v|^2$ term in~\eqref{phi2-ns} diverges.  At spatial infinity $u_{\text{ext}} \to -\infty$.  Using $W(x) \approx \log x$ for $x \gg 1$, one finds that $u_F \approx u_{\text{ext}}$.  Then since $u_{\text{ext}} - v = -2 r_*$, the same term in~\eqref{phi2-ns} diverges at spatial infinity.  On the part of past null infinity that is outside the null shell trajectory, $u_{\text{ext}} \to -\infty$ and $u_F \approx u_{\text{ext}}$ 
so the same term diverges even though $v$ is a regular coordinate on past null infinity.

Past timelike infinity is inside the null shell trajectory.  At future timelike infinity, $u_{\text{ext}} \to +\infty$.  Since $W(x) \approx x$ for $|x| \ll 1$, $u_F \approx v_H - 4 M \exp\left(\frac{v_H - u_{ext}}{4M} \right)$.  In this case, the $-\log|v_H-u_F|$ term in ~\eqref{phi2-ns} is approximately equal to $\frac{v_H - u_{\text{ext}}}{4M} - \log{(4M)}$.  Then for a fixed value of $r$, $\la \phi^2 \ra$ has a linear dependence on $t$ and thus diverges at future timelike infinity.  However, there is a sub-leading term that also diverges there.  It comes from the $\log|v_0-u_F|$ term in~\eqref{phi2-ns} which for fixed $r$ and large $t$ goes like $2 \log t$.  

It turns out that $\la \phi^2 \ra_{in}$ is proportional to $v$ on the future horizon and so diverges there in the limits $v \to \pm \infty$.  To see this note that $u_{ext} \to \infty$ on the future horizon so again $u_F \approx v_H - 4 M \exp\left(\frac{v_H - u_{ext}}{4M} \right)$.
In this case $\frac{v_H - u_{ext}}{4M} - \log(4M) = v_H - v + 2 r_*$.  The divergence in $r_*$ cancels the divergence in the last term of~\eqref{phi2-ns}.  The result is that on the horizon
\be \la \phi^2 \ra_{in} = \frac{1}{4 \pi} \left[2 \gamma_E + \frac{v - v_H}{4 M} - 1 + 2 \log \left( \frac{\mu (v-v_H)}{4} \right) \right] \ee

In the interior, as the singularity is approached, $r_* \to 0$ and it is easy to see that outside of the null shell trajectory the divergence in  $\la \phi^2 \ra_{in}$ is the same as that for $\la \phi^2 \ra_B$.  However, on the null shell trajectory $t = v_0 - r_*$ so $u_{int} = 2 r_* - v_0$ and $u_F(u_{int}) = v_0 -2 r$ where the property $z = W(z e^z)$ has been used.  Then one finds
\be (\la \phi^2 \ra_{in})_{v = v_0} = \frac{1}{4 \pi} (2 \gamma_E + 2 \log\left(\frac{2 r}{M} \right) + \log (m^2 \mu^2) \;. \ee
Thus, there is a discontinuity in $\la \phi^2 \ra_{in}$ at the point where the null shell reaches $r = 0$ as can be seen by the fact that on the null shell trajectory as $r \to 0$, $\la \phi^2 \ra_{\rm in} \sim \frac{1}{2 \pi} \log( r/M)$ while $\la \phi^2 \ra_B \sim - \frac{1}{ 4 \pi} \log(r/M)$.  

\begin{figure}[H]
    \centering
    \includegraphics[width=0.495\linewidth]{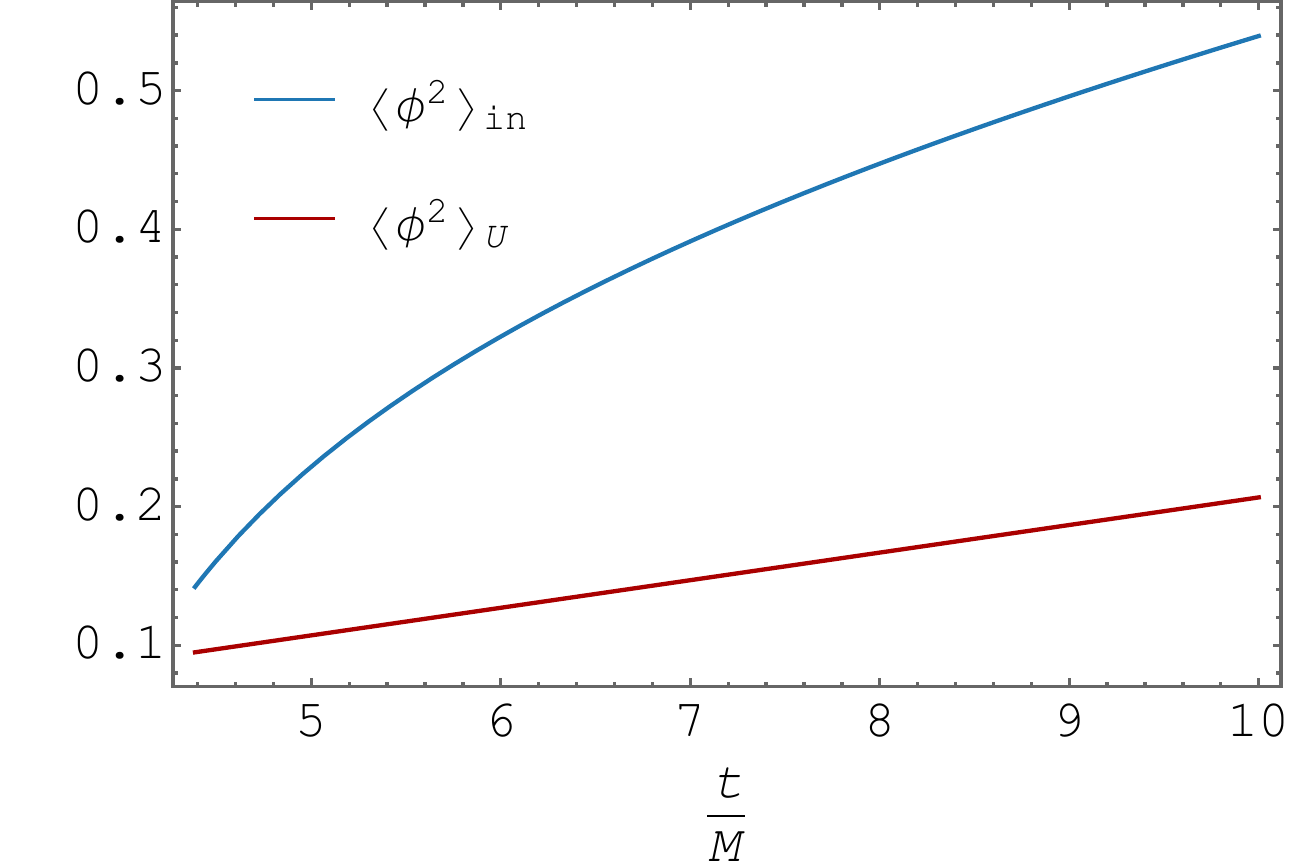}
    \includegraphics[width=0.48\linewidth]{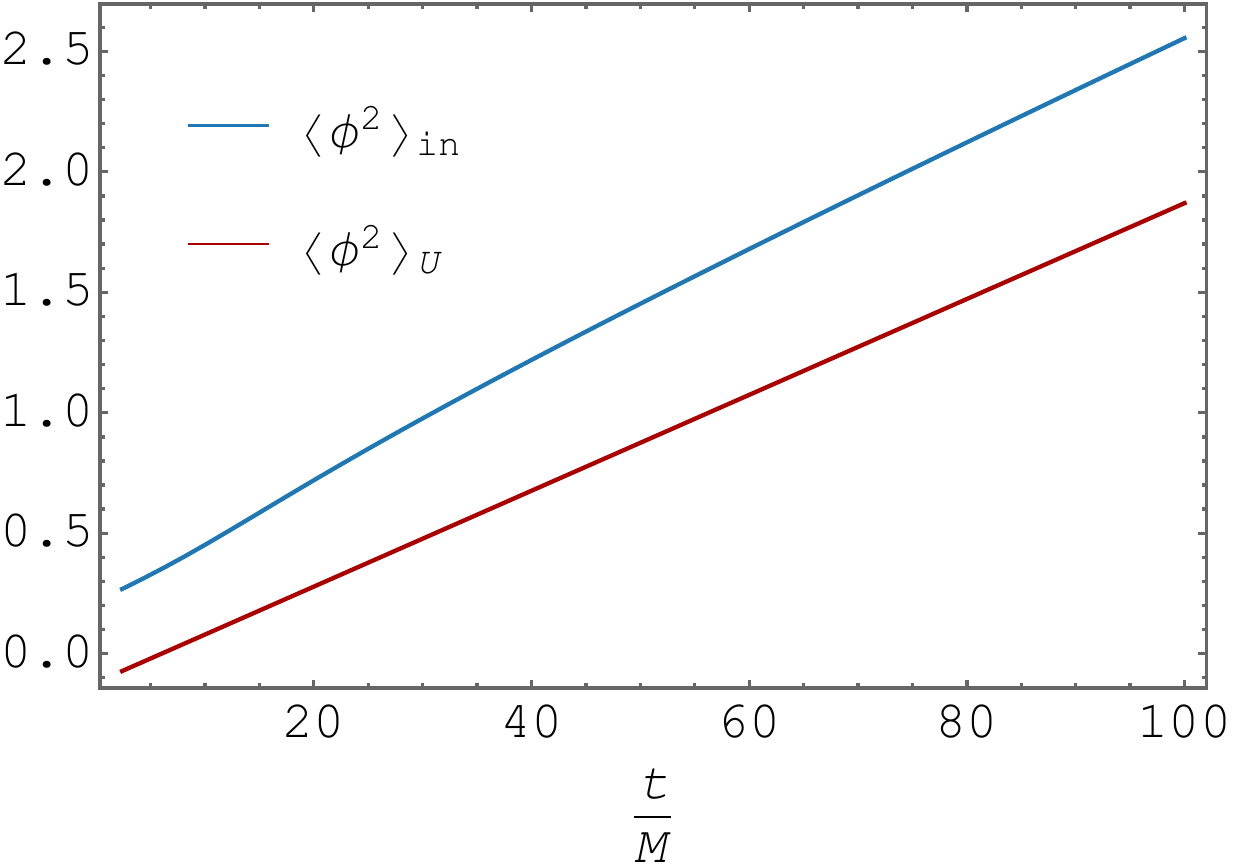}
    \caption{Plots of $\la \phi^2 \ra$ as a function of $\frac{t}{M}$.  The upper (blue) curve corresponds to the {\it in} state with $v_0 = 4$ and the lower (red) curve corresponds to the Unruh state.
   The left panel is for the interior region at $\frac{r}{M} = 1$ and the right panel is for the exterior region at $\frac{r}{M} = 3$. A careful examination of both plots shows that, as expected, $\la \phi^2 \ra_{in}$ is diverging faster than $\la \phi^2 \ra_U$ in both regions. }
    \label{fig:phi2-in}
\end{figure}

For fixed $r$ and large positive values of $t$, $u_{int} = r_* - t \to -\infty $ and $u_F \to v_H + 4M \exp\left(\frac{u_{int}+v+H}{4M} \right)$.  Then $\log(v_H-u_F(u_{int})) \to \frac{u_{int}}{4 M} = \frac{r_* -t}{4M}$ and $\log|v-u_F(u_{int})|^2 \to \log t $. Thus 
$\la \phi^2 \ra_{in}$ has a term that has the same linear dependence on $t$ as $\la \phi^2 \ra_U$.  In addition it has a term that grows logarithmically with $t$. This is illustrated for both the interior and exterior regions in Fig.~\ref{fig:phi2-in}.  Note that the term with the linear growth in $t$ of the two-point function which leads to that in $\la \phi^2 \ra$ disappears when two derivatives of the two-point function are taken to compute the stress-energy tensor.  As a result the stress-energy is independent of $t$ for the Unruh state.  As can be seen from~\eqref{Tab} the stress-energy tensor for the null shell spacetime has no power law or logarithmic behavior in the coordinate $t$.  Thus the $\log t/M$ contribution to $\la \phi^2 \ra_{\rm in}$ does not affect the behavior of the stress-energy tensor.

\section{Stress-Energy Tensor}

As discussed in the Introduction, the stress-energy tensor for the massless minimally-coupled scalar field has been computed previously for the states we are considering.  The focus of previous analyses was the exterior region.  In the appendix, it is shown explicitly that the expressions obtained can be continued into the region inside the future horizon.  In this section, we review their behaviors on the future horizon and analyze them in the interior region.

  The stress-energy tensors can be written in terms of null coordinate components with the result~\cite{Sandro-Book}
  \bes \bea \la T_{uu} \ra_B &=& \frac{1}{24 \pi} \left[ -\frac{M}{r^3} + \frac{3}{2} \frac{M^2}{r^4} \right] \;,  \\
  \la T_{uu} \ra_H &=&  \la T_{uu} \ra_U  = \frac{1}{768 \pi M^2} \left(1 - \frac{2M}{r} \right)^2 \left[1+ \frac{4 M}{r} + \frac{12 M^2}{r^2} \right] \nonumber  \\
   &=& \frac{1}{768 \pi M^2} - \frac{M}{24 \pi r^3} + \frac{M^2}{16 \pi r^4} \;, \\
   \la T_{uu} \ra_{in} &=& \frac{1}{24 \pi} \left[-\frac{M}{r^3} + \frac{3}{2} \frac{M^2}{r^4} - \frac{8 M}{(u_F(u_i) - v_0)^3} - \frac{24 M^2}{(u_F(u_i)-v_0)^4} \right] \;, \label{ns-Tuu} \\  
   \la T_{vv} \ra_B &=& \la T_{uu} \ra_B \;, \\
    \la T_{vv} \ra_H &=& \la T_{uu} \ra_H \;, \\
     \la T_{vv} \ra_U &=& \la T_{vv} \ra_{in} = \la T_{vv} \ra_B \;, \\
         \la T_{uv} \ra_B &=& -\frac{1}{24 \pi} \left(1-\frac{2M}{r} \right) \frac{M}{r^3} \;, 
         \\
     \la T_{uv} \ra_H &=& \la T_{uv} \ra_U = \la T_{uv} \ra_{in} = \la T_{uv} \ra_B
     \; . \eea \label{Tab} \ees  
 
 All of the above components for each of the four states diverges like $r^{-4}$ as the singularity at $r = 0$ is approached.  The coefficients differ in magnitude but they all have the same positive sign. 
 
 As discussed in~\cite{Christensen-Fulling, Sandro-Book}, for the stress-energy to be regular on the event horizon one needs to go to a coordinate frame that is regular there.  For Kruskal coordinates, one finds that for the stress-energy to be finite on the future horizon, 
 $ (r-2M)^{-2} |\la T_{uu} \ra|$, $|(r-2M)^{-1} \la T_{uv} \ra|$, and  $|\la T_{vv} \ra|$ should all be finite there. This is easily seen to be true for the Hartle-Hawking and Unruh states.  As is well known, the stress-energy tensor for the Boulware state diverges on both the past and future horizons. For the {\it in} state, one can first make the substitution in~\eqref{ns-Tuu} 
 \be u_{int} = r_* -t = 2 r_* - v \; \ee
 and note that $v$ is continuous across the future horizon.  Then, using a Taylor series expansion around $r = 2M$ with $v$ held constant, one finds
\be \la T_{uu} \ra_{in} \to (2 M -r)^2 \frac{1}{512 M^4 \pi} \left[-1 + \exp\left(2+\frac{v_H-v}{2 M} \right) \right] \;. \ee
Hence, the stress-energy for the {\it in} state is finite at the horizon.

The flux of energy through the horizon is given by $\la T_{vv} \ra$. It is zero for the Hartle-Hawking state as expected.  For the Boulware, Unruh, and {\it in} states it is $\la T_{vv} \ra = - \frac{1}{768 \pi M^2}$. As is well known, this is equal in magnitude and opposite in sign to the flux of energy $\la T_{uu} \ra$ reaching future null infinity at all times for the Unruh state and at late times for the {\it in} state.  However, it is interesting to note that the same flux of negative energy goes through the future horizon in the Boulware state.  It is also interesting to note that, even though $\la T_{uu} \ra_{\rm in}$ only approaches the value for the Unruh state asymptotically, $\la T_{vv} \ra_{\rm in}$ has the same value as for the Unruh state immediately after the null shell passes through the future horizon.  There are no Einstein equations in 2D, however, if one had similar behavior in 4D, then one would expect that the spacetime geometry near the horizon, after the shell passes through it, would start to evolve in the same way that it would for the Unruh state.  This is very different from how backreaction effects would manifest at large distances from the black hole.   Of course in 4D there is an effective potential for the mode equation of the scalar field which would alter this picture due to backscattering of the outgoing modes through the future horizon.  Nevertheless, one might expect that the flux through the horizon at early times after the null shell passes through it would be very different from the early-time flux going out to future null infinity.

\begin{figure}[H]
    \centering
    \includegraphics[width=0.6\linewidth]{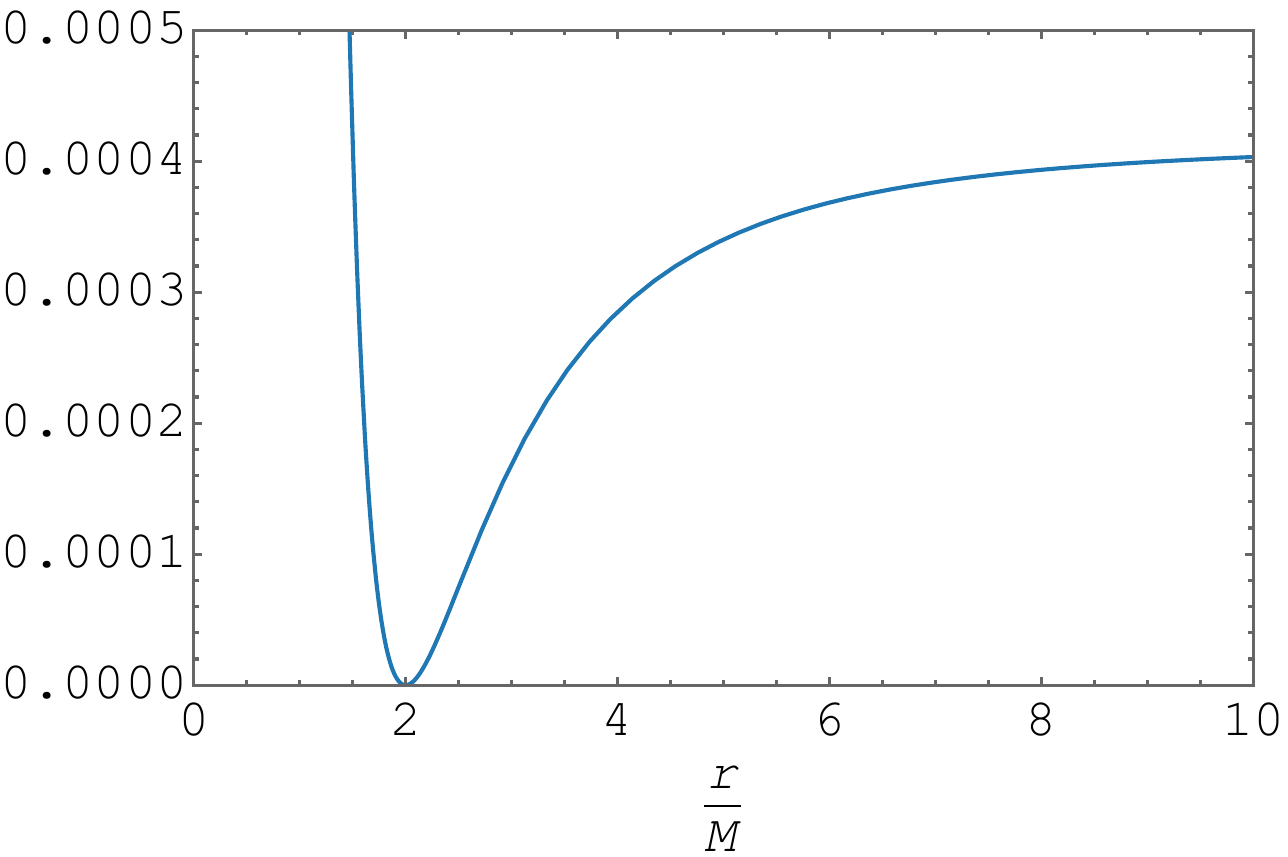}
    \caption{Plot of $M^2 \la T_{uu} \ra_U$ as a function of $\frac{r}{M}$.  The plot covers both the interior and exterior regions. Note that  $\la T_{uu} \ra_U$ diverges at $r = 0$, vanishes at $r = 2M$, and approaches a constant in the limit $r \to \infty$.}
    \label{fig:TuuU}
\end{figure}

 The stress-energy tensor for the null shell spacetime is the only one of the four that we are considering which has a dependence on the coordinate $t$. 
The only component shown in~\eqref{Tab} that is different from that of the Unruh state is $\la T_{uu} \ra_{in}$. Since we are going to compare $\la T_{uu} \ra_{in}$ with $\la T_{uu} \ra_U$,
it is useful to plot the latter as a function of $r$.  This is done in Fig.~\ref{fig:TuuU}.

As was done for $\la \phi^2 \ra_{in}$, using the identity for the Lambert W function, $z = W[z e^z]$, and the fact that on the null shell trajectory, $t = v_0 - r_*$, it is not hard to show that $\la T_{uu} \ra_{in} = 0$ everywhere on the surface $v = v_0$.  Therefore, it is discontinuous at the point where the null shell reaches $r = 0$.  

In the exterior region one expects $\la T_{uu}\ra_U$ to be a good approximation to $\la T_{uu} \ra_{in}$ at late times and this is what is found~\cite{Sandro-Book}.  However, it is not as clear what one would expect inside the future horizon, particularly since $t$ is a space coordinate there.

Before exploring the details of how the Unruh state is approached, it is instructive to examine the behavior of this state which is shown in Fig.~\ref{fig:TuuU}.  There, one sees that this component decreases monotonically from its infinite value at $r = 0$ to zero at the future horizon.  From there, it increases monotonically and approaches a constant value in the limit $r \to \infty$.

\begin{figure}[H]
    \centering
    \includegraphics[width=0.49\linewidth]{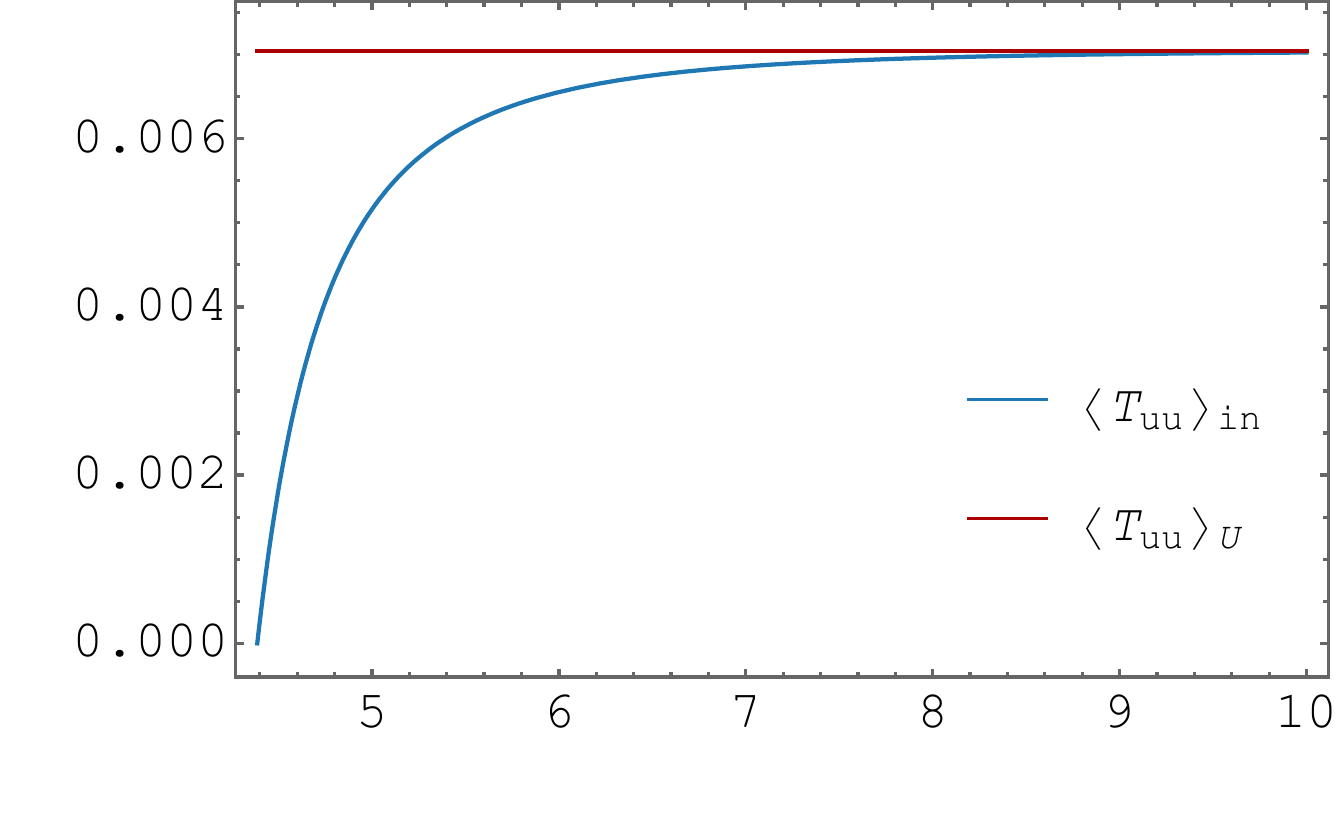}
     \includegraphics[width=0.50\linewidth]{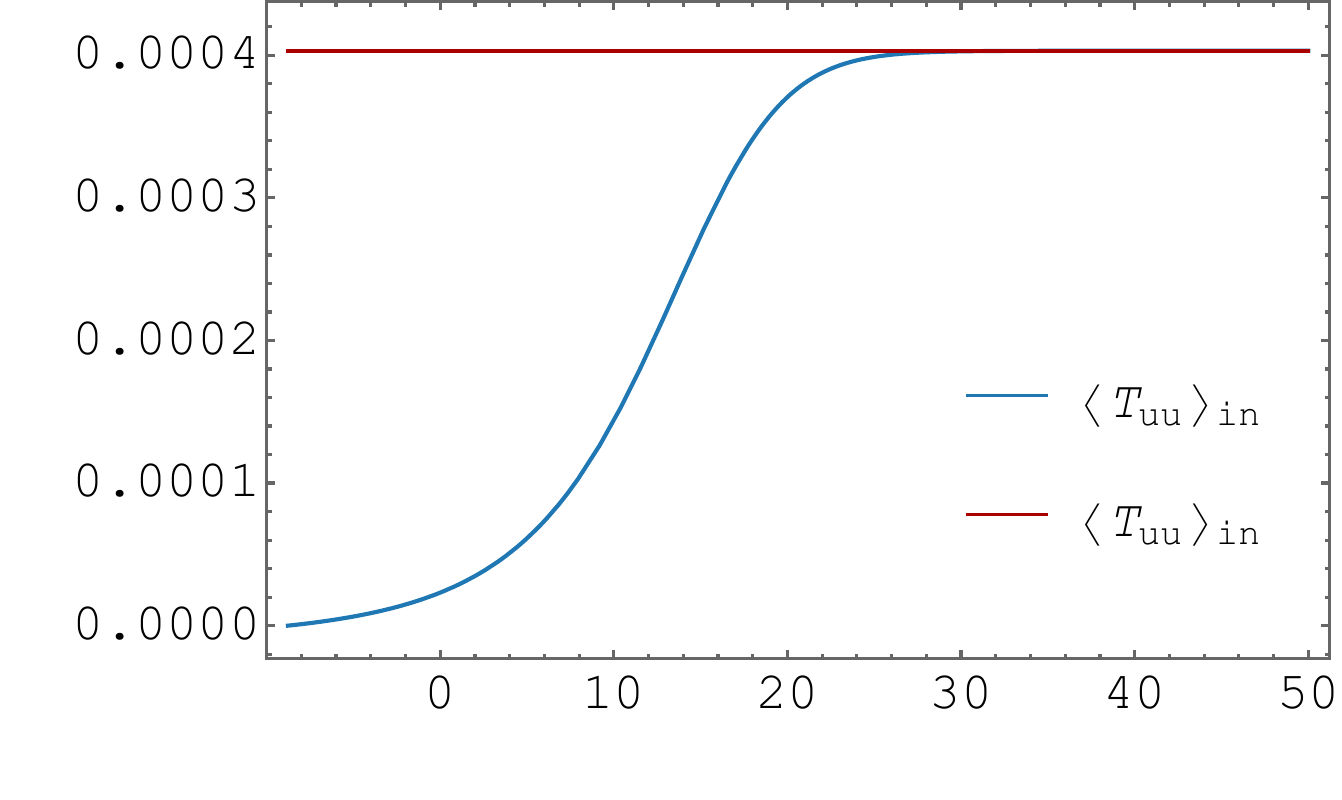}
    \caption{  $M^2 \la T_{uu} \ra_{in}$ is plotted as a function of $\frac{t}{M}$.  For the panel on the left, this is done for the interior region for $\frac{r}{M} = 1$ and for the panel on the right, it is done for the exterior region for $\frac{r}{M} = 3$.  The straight line on both plots is the value of $M^2 \la T_{uu} \ra_U$. }
    \label{fig:Tuuns1}
\end{figure}

Some details as to how the Unruh state is approached are shown in Fig.~\ref{fig:Tuuns1}. Note that the plots begin at the value of $t$ for which the null shell passes through that value of $r$. For a fixed value of $r$ in either the interior region or the exterior region, $\la T_{uu} \ra_{in}$ approaches $\la T_{uu} \ra_U$ at large values of $t$.  The smaller the value of $r$, the more ``quickly'' the approach occurs in both regions.  In the exterior region, this means that the farther away one is from the horizon, the longer it takes for the stress-energy to approach that of the Unruh state.  In the interior region, it means that the closer one is in time ($r$) to the future horizon, the larger the region (in terms of the coordinate $t$) where the stress-energy is significantly different from that for the Unruh state.  

\begin{figure}[H]
    \centering
  \includegraphics[width=0.48\linewidth]{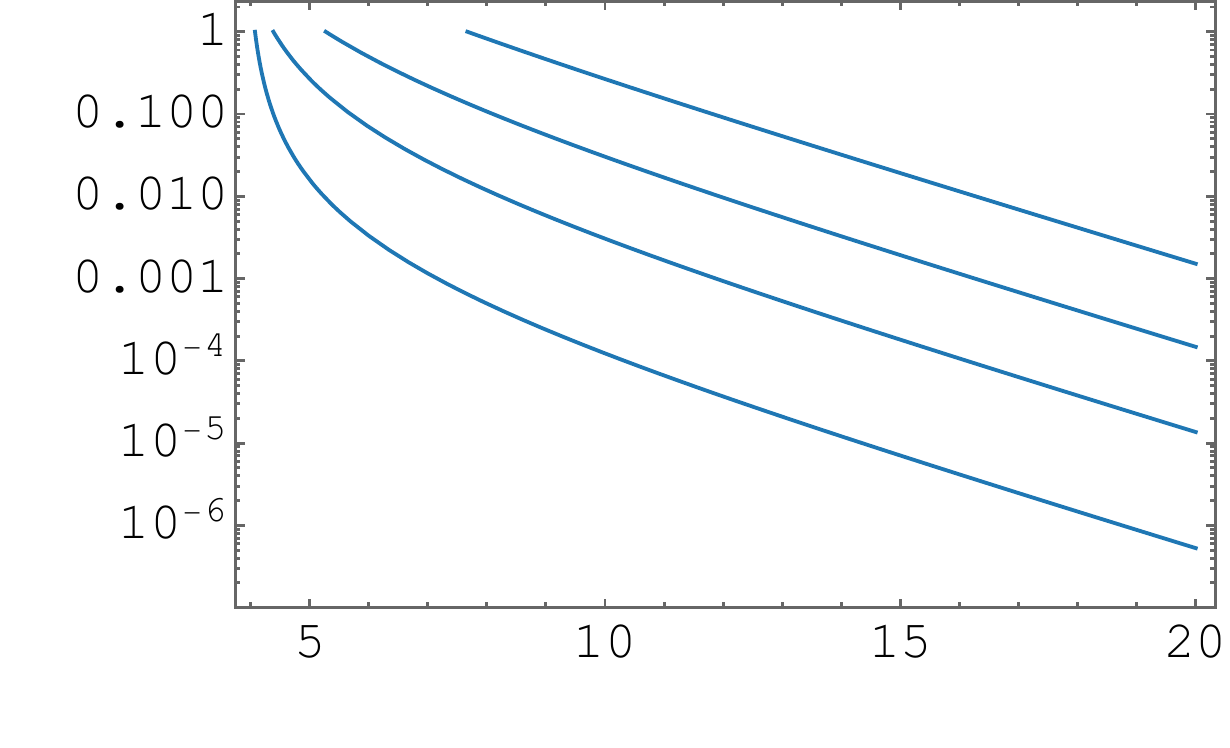}
\includegraphics[width=0.48\linewidth]{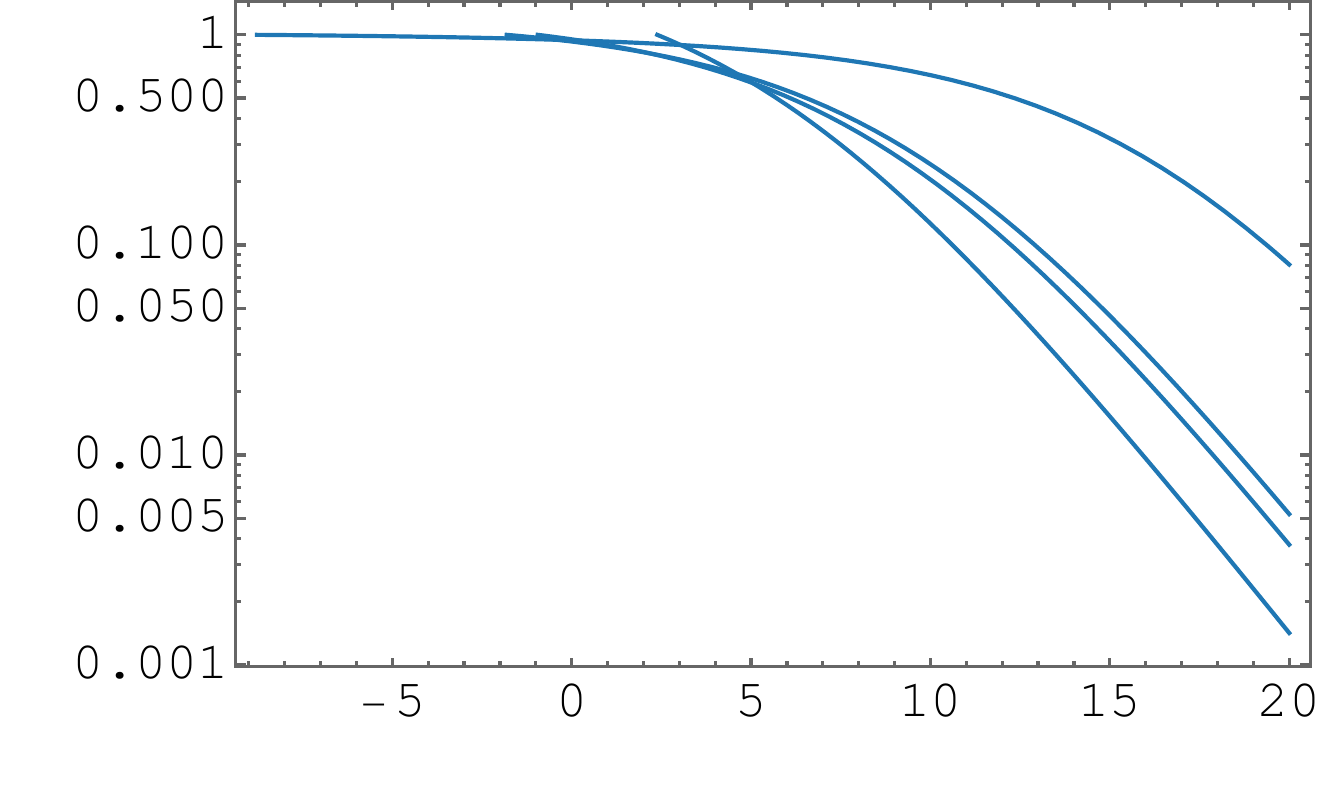}  
  \caption{The absolute value of the relative difference between $\la T_{uu} \ra_{in}$ and $\la T_{uu} \ra_U$ is shown as a function of $\frac{t}{M}$.  For the left panel, the curves from upper to lower correspond to $\frac{r}{M} = \frac{15}{8}$, $\frac{3}{2}$, $1$, and $\frac{1}{2}$.  For the right panel, the curves from upper to lower on the right correspond to $\frac{r}{M} = 10$, $5$, $\frac{9}{2}$, and $3$.  Note that the curve for a given value of $r$, say $r = r_1$, begins at time $t_1 = v_0 - r_*(r_1)$, which is the time that the null shell reaches $r_1$.  The smaller the value of $r_1$, the larger the value of $t_1$ in both the interior and exterior regions.   }
    \label{fig:Tuuns2}
\end{figure}

\section{Conclusions}

We have computed the quantity $\la \phi^2 \ra$ for a massless minimally-coupled scalar field in Schwarzschild spacetime in 2D for the Boulware, Unruh, and Hartle-Hawking states.  We have also computed it for the {\it in} state when the field is in a 2D spacetime where a collapsing null shell forms a black hole.   We have also shown by explicit calculation, that for all of these states, previous calculations of the stress-energy tensor in the region outside the future horizon can be extendable to the region inside the future horizon.

The same type of late-time growth in $\la \phi^2 \ra$ for the Unruh and {\it in} states was found in the exterior region as was previously found for the symmetric two-point function when the points are split in the radial direction~\cite{A-T}. This growth is linear for both states and there is also  logarithmic growth for the {\it in} state.  This is important because it shows that the instability is not an artifact of the point separation.  Similar growth in $t$ occurs for fixed $r$ and large values of $t$ in the interior region.

For the Unruh state, $\la \phi^2 \ra$ is also proportional to $t$ at large negative values of $t$ in both the interior and exterior regions.  This is different from the symmetric two-point function.  As $t \to -\infty$, one sees from~\eqref{Kruskal-U} that the Kruskal coordinates in the interior and exterior regions approach $\pm \infty$ respectively.  Examination of~\eqref{G1-U} with the points split in the radial direction, shows that $G^{(1)}(x,x') \to -\frac{1}{2 \pi} {\rm ci}(r_*-r^\prime_*)$.  Note that this limit does not exist for the null shell spacetime in either region.
This result illustrates the fact that the two-point correlation function can give a very different answer in some cases than the quantity $\la \phi^2 \ra$.

An interesting instability was found for the Hartle-Hawking state due to the divergence of $\la \phi^2 \ra$ in the limit $r \to \infty$.  This includes past and future null infinity and spacelike infinity.  Divergences were also found at all of these locations for the {\it in} state.  

At the future horizon, $\la \phi^2 \ra$ diverges for the Boulware state and is finite for the other states for finite values of $v$ as expected.  However, it does diverge in the limits $v \to \pm \infty$ for the Unruh state and in the limit $v \to + \infty$ for the {\it in} state which is a further indication of the instability of these states in 2D.

In the interior region, $\la \phi^2 \ra$ diverges for all four states exactly in the same way as the singularity at $r = 0$ is approached.  For the Unruh state, there is a term proportional to $t$ which is a spatial coordinate in the interior.  For the {\it in} state, the dependence on $t$ is more complicated.  At fixed values of $r$ there is both a term proportional to $t$ and one proportional to $\log t$ at large values of $t$. The dependence on $t$ indicates that the Unruh and {\it in} states are inhomogeneous in the interior.

The instabilities in $\la \phi^2 \ra$ mentioned above in both the interior and exterior regions are not reflected in the 2D stress-energy tensors for the field.  These stress-energy tensors have been previously studied in the exterior region and at the future horizon~\cite{Sandro-Book}.  

The stress-energy for all four states diverges at the singularity as expected.  As is well known, on the future horizon the only divergence in the stress-energy is for the Boulware state.

The only state for which the stress-energy has any dependence on $t$ is the {\it in} state.  In terms of null components, only $\la T_{uu}\ra$ has a dependence on $t$.  This component vanishes on the null shell trajectory and increases outside of it.  At fixed $r$ for large values of $t$ in both the interior and exterior regions, $\la T_{uu} \ra_{in}$ approaches $\la T_{uu} \ra_U$.  

The dependence of the stress-energy on $t$ in the interior region will occur for any black hole that forms from collapse in 2D or 4D.  An important question is how well the Unruh state approximates quantum effects in the interior for a black hole in this case.  This is because it is much easier in 4D to compute the stress-energy tensor for the Unruh state in the interior of an eternal black hole than it would be to compute it for one that forms from collapse.  We find that, for fixed values of $r$ as $r$ gets smaller the range of values of $t$ for which there is a significant difference from the stress tensor for the Unruh state decreases.  At the same time, the spacetime curvature and the stress-energy increase.  If something similar happens in 4D, then backreaction effects might be reasonably well approximated throughout much of the spacetime outside the collapsing matter by considering the Unruh state. 

\acknowledgements

We would like to thank Alessandro Fabbri for helpful comments about the manuscript.  This work was supported in part by the National Science Foundation under Grants No. PHY-1505875,  PHY-1912584 and PHY-2309186 to Wake Forest University.

\newpage
\appendix
\counterwithin*{equation}{section}
\renewcommand\theequation{\thesection\arabic{equation}}
\section{Stress-energy tensor for a 2D Schwarzschild black hole }
\label{appendix1}
In this appendix, we derive an expression for the renormalized stress-energy tensor for a massless minimally-coupled scalar field of a 2D Schwarzschild black hole that works for the Boulware, Unruh, Hartle-Hawking, and {\it in} states and is valid in both the interior and exterior regions. We use the covariant point splitting method in~\cite{Bunch-Christensen-Fulling}.  This is equivalent to previous derivations using the Schwarzian derivative~\cite{Sandro-Book}.  We explicitly show that the expressions for the stress-energy tensors for these states that were previously derived work for both the interior and exterior regions.  At the end we also discuss the point splitting counter terms for $\la \phi^2 \ra$.

A general expression for the unrenormalized  stress-energy tensor for a massless minimally-coupled scalar field and with the points split is of form
\bea
\langle T_{\mu \nu}\rangle_{\text{unren}}=\frac{1}{4}\Big\{g_{\mu}^{\alpha'}G_{;\alpha' \nu}^{(1)}+g_{\nu}^{\alpha'}G_{\;\mu \alpha'}^{(1)}-g_{\mu \nu}g^{\sigma \alpha'}G^{(1)}_{\sigma \alpha'}\Big\},\label{app-tunren}
\eea
where $G^{(1)}(x,x')$ is the symmetric two-point correlation function also called the Hadamard function.  The bivector of parallel transport, $g_\mu^{\alpha'}$, is used to parallel transport vectors and tensors with primed indices from the point $x'$ to the point $x$~\cite{Christensen1976}.

To go further, it is easiest to write the metric in the form
\be ds^2 = -f dt^2 + \frac{dr^2}{f} \ee
with $f = 1 - \frac{2M}{r}$.  The main difference with the exterior region is that in the interior $f < 0$ so $t$ is a space coordinate and $r$ is a time coordinate.  
To obtain an explicit expression, we separate the points $x$ and $x'$ in the $t$ direction such that $x=(t,r)$ and $x'=(t',r)$ with $t' = t + \epsilon$.  To renormalize the stress-energy tensor, we first expand $\langle T_{\mu \nu}\rangle_{\text{unren}}$ in powers of $\epsilon$ and then expand the covariant point splitting counterterms in powers of $\epsilon$.  Taking the difference between the two expressions and setting $\epsilon = 0$ gives the renormalized expression for the stress-energy tensor.

To carry out this procedure, it is first necessary to consider derivatives of the quantity $\sigma(x,x')$, which is one-half of the square of the proper distance between the points $x$ and $x'$ along the shortest geodesic connecting them.  It satisfies the relation
\be \sigma = \frac{1}{2} \sigma_\mu \sigma^{\mu} \ee
with $\sigma_{\mu}$ the covariant derivative with respect to $x^\mu$ and $\sigma^{\mu}$ is the geodesic's tangent vector at $x$ which points towards $x'$. There is no difference in the derivation of its expansion in powers of $\epsilon$ between the interior and exterior regions.  
In \cite{AHS-2}, the following expressions for $\sigma^{t}$ and $\sigma^r$ are given in powers of $\epsilon$
 \bes \bea
\sigma^t&=&\sigma^t(\epsilon,r)=\epsilon+\frac{f'^2}{24}\epsilon^3+\mathcal{O}(\epsilon^5),\\
\sigma^{r}&=&\sigma^{r}(\epsilon,r)=-\frac{f f'}{4}\epsilon^2+\mathcal{O}
(\epsilon^4).  \eea  \label{sigmatr}\ees

The bivector of parallel transport is a vector at both $x$ and $x'$.  Thus it satisfies the relation
$g_{\mu}^{\alpha'}=g_{\mu \rho}g^{\rho \alpha'}$.  Using this, one can write ~\eqref{app-tunren} in the form
\bea
\langle T_{\mu \nu}\rangle_{\text{unren}}=\frac{1}{2}g_{\mu \rho}g^{\rho\alpha'}G_{;\alpha' \nu}^{(1)}-\frac{1}{4}g_{\mu \nu}g^{\sigma \alpha'}G_{\;\sigma \alpha'}^{(1)}.\label{app-tunren2}
\eea
One way to compute $g^{\rho \alpha'}$  is to define two suitable orthonormal frames at points $x$ and $x'$ and use the relations 
\bea
g^{\mu \nu'}=\eta^{ab} e_a^{\mu} \, e_b^{\nu'},\label{bitensor} 
\eea
where $e_a^{\mu}$ and $e_b^{\nu'}$ are the orthonormal basis vectors of tetrads at $x$ and $x'$ respectively. Following the method given in~\cite{Howard,AHS-2} , we set $e_0^\mu$ to be in the time direction and $e_1^\mu$ to be in the space direction.  In the interior, these are in the $r$ and $t$ directions respectively.  The orthonormality relation is
\be \eta_{a b} = g_{\mu \nu} e_a^{\; \mu} \; e_b^{\; \nu} \;, \label{ortho-basis-vectors} \ee
with $\eta_{ab}$ the usual Minkowski metric.
Since the shortest geodesic from the point $(t',r)$ to the point $(t,r)$ is spacelike in the interior and since its tangent vector is $\sigma^\mu$, we set $e_1^\mu = N \sigma^\mu$.  Then~\eqref{ortho-basis-vectors} gives $N = (\sigma_\mu \sigma^\mu) ^{-1/2}$.  Since ~\eqref{ortho-basis-vectors} implies that $e_0^{\; \mu}$ and $e_1^{\; \mu}$ are orthogonal, we can use this property to find $e_0^\mu$.  The result is 
\bes \bea
e_1^{t}&=&\frac{\sigma^{t}}{(\sigma_\alpha \sigma^{\alpha})^{\frac{1}{2}}}, \quad e_1^{r}=\frac{\sigma^{r}}{(\sigma_\alpha \sigma^{\alpha})^{\frac{1}{2}}}, \\
 e_0^t&=& \frac{1}{f}\frac{\sigma^r}{(\sigma_{\alpha} \sigma^{\alpha})^{\frac{1}{2}}}, \quad e_0^r=f\frac{\sigma^t}{(\sigma_{\alpha} \sigma^{\alpha})^{\frac{1}{2}}}.\label{tetrad1}
\eea \ees
Aside from the fact that we are working in 2D, there are two differences between this result and the exterior result derived in~\cite{Howard, AHS-2}.  One is that for this result the normalization is $(\sigma_\mu \sigma^\mu)^{-1/2}$ rather than  $(-\sigma_\mu \sigma^\mu)^{-1/2}$, which is its value in the exterior region.  The other is the switching of the basis vectors so that $e_0^\mu$ in the exterior region goes to $e_1^\mu$ in the interior region and $e_1^\mu$ in the exterior region goes to $e_0^\mu$ in the interior region.  The reason is simply that $t$ is a space coordinate in the interior region and a time coordinate in the exterior one.

One can define a similar set of tetrad basis vectors at $x'$
\bes \bea
e_1^{t'}&=&\frac{\sigma^{t'}}{(\sigma_{\alpha'} \sigma^{\alpha'})^{\frac{1}{2}}}, \quad e_1^{r'}=\frac{\sigma^{r'}}{(\sigma_{\alpha'}\sigma^{\alpha'})^{\frac{1}{2}}}, \\
 e_0^{t'}&=&\frac{1}{f}\frac{\sigma^{r'}}{(\sigma_{\alpha'} \sigma^{\alpha'})^{\frac{1}{2}}}, \quad e_0^{r'}= f\frac{\sigma^{t'}}{(\sigma_{\alpha'} \sigma^{\alpha'})^{\frac{1}{2}}}.\label{tetrad2}
\eea \ees
It is now necessary to expand $\sigma^{\mu'}$ as a power series in $\epsilon'=-\epsilon$.  The method is the same as for the expansion of $\sigma^\mu$ so one can simply substitute $\epsilon'$ for $\epsilon$ in~\eqref{sigmatr}, and then let $\epsilon'=-\epsilon$ with the result  
\bes \bea
\sigma^{t'}(\epsilon',r)&=&-\sigma^t(\epsilon,r),\\
\sigma^{r'}(\epsilon',r)&=&\sigma^r(\epsilon,r),\\
 \left[ \sigma_{\alpha'} (\epsilon',r)\sigma^{\alpha'}(\epsilon',r) \right]^{\frac{1}{2}} &=& -\left[\sigma_{\alpha}(\epsilon,r)\sigma^{\alpha}(\epsilon,r) \right]^{\frac{1}{2}}. 
\eea \ees
Note that the expansion for  $[\sigma_{\alpha'}(\epsilon',r)\sigma^{\alpha'}(\epsilon',r)]^{\frac{1}{2}}$ ends up being in odd powers of $\epsilon'$ which is why it has the opposite sign as $[\sigma_{\alpha}(\epsilon,r)\sigma^{\alpha}(\epsilon,r)]^{\frac{1}{2}}$ .

Therefore, the components of the bi-tensor parallel transport  take the following form
\bes \bea
g^{tt'}&=& \frac{1}{\sigma_{\alpha}\sigma^{\alpha}}\Big[\frac{1}{f^2}(\sigma^r)^2 + (\sigma^t)^2\Big]
 = -\frac{1}{f} - \frac{f^{\prime\, 2}}{8 f} \epsilon^2 + O(\epsilon^4) \;,\\
g^{tr'} &=& - g^{r t'} =  -\frac{2 \sigma^r \sigma^t}{\sigma_{\alpha}\sigma^{\alpha}} = -\frac{f^\prime}{2} \epsilon + O(\epsilon^3) \;,  \\
g^{rr'}&=& -\frac{1}{\sigma_{\alpha}\sigma^{\alpha}}\Big[f^2 (\sigma^t)^2+(\sigma^r)^2\Big] =
f + \frac{f f^{\prime \, 2}}{8} \epsilon^2 + O(\epsilon^4) \;.
\eea \label{bitensor3}\ees 
The result in powers of $\epsilon$ is exactly the same as for the exterior region given in~\cite{AHS-2}.

For all four states we can write the symmetric two-point function in the general form
\be G^{(1)}(x,x') =  -\frac{1}{2 \pi} \{{\rm ci}[\lambda (\mathscr{U}(u_i)-\mathscr{U}(u'_i))] + {\rm ci}[\lambda (\mathscr{V}(v)-\mathscr{V}(v')]\} \;. \label{G1-gen} \ee
For the Boulware state $U(u_i) = u_i$ and $V(v) = v$, for the Hartle-Hawking state $\mathscr{U} = U$ and $\mathscr{V} = V$, and for the {\it in} state $\mathscr{U} = u_F$.
We next give the general expansion in powers of $\epsilon$ for two derivatives of $G$.  Note that $\frac{d}{dz} {\rm ci}(\lambda z) = \frac{\cos(\lambda z)}{z}$.  Therefore after computing the first derivative we can set $\lambda = 0$ since there is no longer an infrared divergence.  The derivatives of $G^{(1)}$ that contribute to the stress-energy tensor are
\bes \bea G^{(1)}_{,u_i u^\prime_i}(x,x')&=&-\frac{1}{2\pi} \frac{\frac{d \mathscr{U}(u_i)}{d u_i} \frac{d \mathscr{U}(u^\prime_i)}{d u^\prime_i}}{(\mathscr{U}(u_i)-\mathscr{U}(u'_i))^2}  \;, \\
G^{(1)}_{,v v^\prime_i}(x,x')&=&-\frac{1}{2\pi} \frac{\frac{d \mathscr{V}(v)}{d v} \frac{d \mathscr{V}(v^\prime)}{d v^\prime}}{(\mathscr{V}(v-\mathscr{V}(v))^2}.
\eea \label{G-derivs}  \ees 
Next, one can expand $\mathscr{U}(u^\prime_i)$ and $\mathscr{V}(v^\prime)$ in power series with the result
\bes \bea  \mathscr{U}(u^\prime_i) &=& \mathscr{U}(u_i) - \left(\frac{d\mathscr{U}(u_i)}{d u_i} \right) (u_i - u^\prime_i) + \ldots \;, \\
\mathscr{V}(v^\prime) &=& \mathscr{V}(v) - \left(\frac{d\mathscr{V}(v)}{d v} \right) (v - v^\prime) + \ldots\;.  \eea \ees
With the definition
\be \epsilon = t - t'  \;, \ee
one  easily finds the relations
\bea v - v^\prime &=& \epsilon \;, \nonumber \\
u_{\rm int} - u^\prime_{\rm int} &=& - \epsilon \;, \nonumber \\
u_{\rm ext} - u^\prime_{\rm ext} &=& \epsilon \;. \eea
Then one finds  
 \bea G^{(1)}_{,u_i u^\prime_i}(x,x')&=& -\frac{1}{2 \pi} \left[\frac{1}{\epsilon^2} - \frac{1}{4} \left(\frac{ \frac{d^2\mathscr{U}(u_i)}{d u^2_i} }{ \frac{d\mathscr{U}(u_i)}{d u_i}} \right)^2
+ \frac{1}{6} \frac{\frac{d^3\mathscr{U}(u_i)}{d u^3_i}} {\frac{d\mathscr{U}(u_i)}{d u_i}} \right] \label{GD1}\;, \\  G^{(1)}_{,v v^\prime}(x,x')&=& -\frac{1}{2 \pi} \left[\frac{1}{\epsilon^2} - \frac{1}{4} \left(\frac{ \frac{d^2\mathscr{V}(v)}{d v^2} }{ \frac{d\mathscr{V}(v)}{d v}} \right)^2
+ \frac{1}{6} \frac{\frac{d^3\mathscr{V}(v)}{d v^3}} {\frac{d\mathscr{V}(v)}{d v}} \right] \label{GD2}\;.
\eea 
For the Hartle-Hawking state $\mathscr{V}(v) = V(v) = 4M e^{\frac{v}{4M}}$ in both the interior and exterior regions.  For the other three states, $\mathscr{V}(v) = v$.  Thus, $G^{(1)}_{,v v^\prime}(x,x')$ is the same function of $v$ in both regions.  For the Boulware state $\mathscr{U}(u_i) = u_i$ so $G^{(1)}_{,u_i u^\prime_i}(x,x')$ is the same in both regions.  For the other states it is necessary to compute $G^{(1)}_{,u_i u^\prime_i}(x,x')$ separately in both regions. 
The results can be written in the general form
\bea G^{(1)}_{,u_i u^\prime_i}(x,x')&=& -\frac{1}{2 \pi \epsilon^2} + \mathscr{A} \;, 
\nonumber \\  G^{(1)}_{,v v\prime}(x,x')&=& -\frac{1}{2 \pi \epsilon^2}  + \mathscr{B} \;, \eea
Where $\mathscr{A}$ and $\mathscr{B}$ are state-dependent constants.
For the states we consider and the notation that we use, the values of $\mathscr{A}$ and $\mathscr{B}$ for a given state are the same for both the interior and exterior regions.  For the Boulware state
\be \mathscr{A}_B = \mathscr{B} = 0 \;, \ee
for the Unruh state
\bea \mathscr{A}_U &=& \frac{1}{384 \pi M^2}  \qquad \mathscr{B}_U = 0 \;, \eea
for the Hartle-Hawking state
\be \mathscr{A}_H = \mathscr{B}_H = \frac{1}{384 \pi M^2}   \;, \ee
and for the {\it in} state  

\bea \mathscr{A}_{\rm in} &=& -\frac{2 M}{3 \pi  [u_F(u_i) - v_0]^3} - \frac{2 M^2}{\pi [u_F(u_i)-v_0]^4} 
\;, \nonumber \\
 \mathscr{B}_{\rm in} &=& 0 \;, \eea

When the points are split in the $t$ direction, it is easy to show using the chain rule and the definitions of $u_i$ and $v$ that
\bes \bea
G^{(1)}_{,t t'}(x,x') & = &   G^{(1)}_{,u_i u^\prime_i}(x,x') + G^{(1)}_{,v v'}(x,x') \;, \\
G^{(1)}_{,t r'}(x,x') & = &  G^{(1)}_{,r t'}(x,x') = \frac{1}{f} [ - G^{(1)}_{,u_i u^\prime_i}(x,x') + G^{(1)}_{,v v'}(x,x') ] \;, \\
G^{(1)}_{,r r'}(x,x') & = &  \frac{1}{f^2} [G^{(1)}_{,u_i u^\prime_i}(x,x') + G^{(1)}_{,v v'}(x,x') ]
 =  \frac{1}{f^2} G^{(1)}_{,t t'}(x,x') \;, \eea \label{GDtr} \ees
where $f = 1 - \frac{2M}{r}$.
By substitution of  ~\eqref{GDtr} and ~\eqref{bitensor3} into ~\eqref{app-tunren2}, we find 
\bes \bea
\langle T_{tt}\rangle_{\text{unren}}&=& f^2 \langle T_{rr}\rangle_{\text{unren}} = -\frac{1}{2\pi \epsilon^2}-\frac{f'^2}{16\pi f h}  + \frac{1}{2} (\mathscr{A}+\mathscr{B}) \;,\nonumber \\
 \label{Tttunren}\\
\langle T_{tr}\rangle_{\text{unren}}&=&-\frac{f'}{4\pi f \epsilon} - \frac{1}{2 f} (\mathscr{A} - \mathscr{B}) + O(\epsilon) \;. \label{Ttrunren}
\eea \label{Tabunren} \ees

To renormalize   $\langle T_{\mu \nu}\rangle_{\text{unren}}$,  we subtract the renormalization counter-terms and let the points come together. The renormalization counter-terms are computed from the DeWitt-Schwinger expansion of the symmetric two-point function.  The result in 2D is~\cite{Bunch-Christensen-Fulling}
\bea
\langle T_{\mu \nu}(x,x')\rangle_{\text{DS}}=\frac{1}{2\pi}\Big\{ (\sigma_{\beta}\sigma^{\beta})^{-1}\big(g_{\mu \nu}-2\frac{\sigma_{\mu}\sigma_{\nu}}{\sigma_{\alpha}\sigma^{\alpha}}\big)-\frac{1}{12}R\frac{\sigma_{\mu}\sigma_{\nu}}{\sigma_{\alpha}\sigma^{\alpha}}\Big\}.\label{TabDS}
\eea

with $R=-f''$.
Substituting~\eqref{sigmatr} into~\eqref{TabDS} and subtracting the result from ~\eqref{Tabunren} one finds for the renormalized components of the stress-energy tensor
\bes \bea 
\langle T_{tt}\rangle&=& -\frac{f'^2}{96\pi}+\frac{f f''}{24\pi} + \frac{1}{2} (\mathscr{A} + \mathscr{B}) \;, \\
\langle T_{tr}\rangle&=& -\frac{1}{2 f} (\mathscr{A} - \mathscr{B} ) \;, \\
\langle T_{rr}\rangle&=&  -\frac{f'^2}{96\pi} + \frac{1}{2} (\mathscr{A} + \mathscr{B}) \;.
\eea \ees
Next we convert to the null coordinate components. to obtain
\bes \bea 
\la T_{uu} \ra &=& \frac{1}{4} ( \la T_{tt} \ra + f^2 \la T_{rr} \ra) - \frac{1}{2} f \la T_{tr} \ra = -\frac{f'^2}{192\pi} +\frac{f f''}{96\pi} + \frac{\mathscr{A}}{2}   \;, \\
\la T_{uv} \ra &=& \frac{1}{4} ( \la T_{tt} \ra - f^2 \la T_{rr} \ra) =  \frac{f f''}{96 \pi}  \;, \\
\la T_{vv} \ra &=& \frac{1}{4} ( \la T_{tt} \ra + f^2 \la T_{rr} \ra) + \frac{1}{2} f \la T_{tr} \ra =
-\frac{f'^2}{192\pi} +\frac{f f''}{96\pi} + \frac{\mathscr{B}}{2} \;. 
\eea \ees 
Evaluating these in each of the four states gives the expressions in~\eqref{Tab} which we have now shown to be valid in both the interior and exterior regions.

The quantity $\la \phi^2 \ra$ also needs to be renormalized.  For a massless minimally-coupled scalar field, the renormalization counter term in a general spacetime is given by ~\cite{Bunch-Christensen-Fulling}
\be \la \phi^2 \ra_{ps} = \frac{1}{2} G^{(1)}(x,x')_{DS} = -\frac{1}{2\pi}\bigg[\gamma +\frac{1}{2}\ln{\bigg(\frac{\mu^2 |\sigma|}{2}\bigg)}\bigg]+\mathcal{O}(\sigma).
\ee
Using the expansion in~\eqref{sigmatr} along with the relation $\sigma = \frac{1}{2} \sigma_\alpha\sigma^\alpha$ one finds the expression in~\eqref{phi2-ps}.


\begin{thebibliography}{99}
   
     \bibitem{Boulware}
     D. Boulware. ``Quantum field theory in Schwarzchild and Rindler spaces'' \textit{Phys. Rev. D} \textbf{11} 1404 (1975).

\bibitem{H-H} 
J. B. Hartle and S. W. Hawking, ``Path-integral derivation of black-hole radiance'', 
\textit{Phys. Rev. D} \textbf{13}, 2188 (1976).

  \bibitem{Unruh}
     W. G. Unruh, ``Notes on black-hole evaporation'', \textit{Phys. Rev. D} \textbf{14}, 870 (1976).     
\bibitem{Davies-Fulling-Unruh} P. C. W. Davies, S. A. Fulling, W. G. Unruh, ``Energy-momentum tensor near an evaporating black hole'', \textit{Phys. Rev. D} \textbf{13}, 2720 (1976).

 \bibitem{Fulling}
     S. A. Fulling ``Alternative vacuum states in static space-times with horizon'', \textit{J. Phys. A: Math. Gen.} \textbf{10}, 917 (1977).

 \bibitem{Hiscock}
     W. Hiscock, ``Models of evaporating black holes I'', \textit{Phys. Rev. D} \textbf{23} 2813 (1981).

 \bibitem{Sandro-Book}
     See for example, A. Fabbri and J. Navarro-Salas, ``Modeling black hole evaporation'', Imperial College Press, London (2005) and references contained therein.

\bibitem{Sandro-Roberto} R. Balbinot and A. Fabbri,   ``Quantum correlations across the horizon in acoustic and gravitational Black Holes'',  \textit{Phys. Rev. D} {\bf 105}, 045010 (2022).  


\bibitem{Fawcett-Whiting} M. S. Fawcett and B. F. Whiting, in {\it Quantum Structure of Space and Time}, proceedings of the Nuffield Workshop, Imperial College, London, edited by M. J. Duff and C. J. Isham (Cambridge University Press, Cambridge, England, 1982).

\bibitem{Candelas-Howard} P. Candelas and K. W. Howard, \textit{Phys. Rev. D} \textbf{29}, 1618 (1984).

\bibitem{And-phi2} P. R. Anderson, ``$\la \phi^2 \ra$ for massive fields in Schwarzschild spacetime'', \textit{Phys. Rev. D}{\bf 39}, 3785 (1989).

\bibitem{Levi-Ori-phi2-1} A. Levi and A. Ori, ``Pragmatic mode-sum regularization method for semiclassical black-hole spacetimes'', \textit{Phys. Rev. D} {\bf 91}, 104028 (2015).

\bibitem{Levi-Ori-phi2-2} A. Levi and A. Ori, ``Mode-sum regularization of $\la \phi^2 \ra$ in the angular-splitting method'',  \textit{Phys. Rev. D} \textbf{94}, 044054 (2016).  

\bibitem{Fawcett} M. S. Fawcett, \textit{Comm. Math. Phys.} {\bf 89}, 103 (1983).

\bibitem{Howard-Candelas} K. W. Howard and P. Candelas, \textit{Phys. Rev. Lett.} {\bf 53}, 403 (1984).

\bibitem{Howard}
    K. Howard. `` Vacuum $\langle T_{\mu}^{\nu}\rangle$ in Schwarzschild spacetime'', \textit{Phys. Rev. D} \textbf{30} 2532 (1984). 
    
\bibitem{Jensen-McLaughlin-Ottewill-1}  B. P. Jensen, J. G. Mc Laughlin, and A. C. Ottewill, ``Renormalized electromagnetic stress tensor for an evaporating black hole'', \textit{Phys. Rev. D} {\bf  43}, 4142 (1991).

\bibitem{Jensen-McLaughlin-Ottewill-2} B. P. Jensen, J. G. Mc Laughlin, and A. C. Ottewill, ``Anisotropy of the quantum thermal state in Schwarzschild space-time'', \textit{Phys. Rev. D} {\bf 45}, 3002 (1992).

\bibitem{AHS-1} P. R. Anderson, W. A. Hiscock, and D. A. Samuel, ``Stress-Energy Tensor of Quantized Scalar Fields in Static Black Hole Spacetimes'', \textit{Phys. Rev. Lett.} {\bf 70}, 1739 (1993).

\bibitem{AHS-2} P. R. Anderson, W. A. Hiscock, and D. A. Samuel, ``Stress-energy tensor of quantized scalar fields in static spherically symmetric spacetimes'', \textit{Phys. Rev. D} {\bf 51}, 4337 (1995).

\bibitem{And-Balbinot-Fabbri} P. R. Anderson, R. Balbinot, and A. Fabbri, ``Cutoff Anti–de Sitter Space/Conformal-Field-Theory Duality and the Quest for Braneworld Black Holes'', \textit{Phys. Rev. Lett.}, {\bf 94}, 061301 (2005).

\bibitem{Levi-Ori-Tab} A. Levi and A. Ori, ``Versatile method for renormalized stress-energy computation in black-hole spacetimes'', \textit{Phys. Rev. Lett.} {\bf 117}, 231101 (2016).

\bibitem{Taylor-Breen-Ottewill}  P. Taylor, C. Breen, A. Ottewill, ``A mode-sum prescription for the renormalized stress energy tensor on black hole spacetimes'', \textit{ Physical Review D} \textbf{106}, 065023 (2022).

\bibitem{Candelas-Jensen} P. Candelas and B. P. Jensen, \textit{Phys. Rev. D}{\bf 33}, 1596 (1986).

\bibitem{Lanir-Levi-Ori-phi2-inside} A. Lanir, A. Levi, and A. Ori, ``Mode-sum renormalization of     for a quantum scalar field inside a Schwarzschild black hole'',  \textit{Phys. Rev. D} {\bf 98}, 084017 (2018). 

 \bibitem{Ori-Zilberman}
     A. Ori and N. Zilberman, ``Computation of the semiclassical outflux emerging from a collapsing spherical null shell'' arXiv:2503.00622.
    

  
 \bibitem{A-T}
 P. R. Anderson, J. Traschen, ``Horizons and correlation functions in 2D Schwarzchild-de Sitter spacetime'', \textit{J. High Energy Physics} \textbf{192}, 01 (2022).

\bibitem{A-G-S} P. R. Anderson, S. Gholizadeh Siahmazgi, Z. P. Scofield,   ``Infrared Effects and the Unruh State'', \textit{Class. Quantum Grav.} \textbf{40} 135004.

\bibitem{N-P-A}
I. M. Newsome, S. Pla, P. R. Anderson,``Quantum effects in 3+1 Schwarzschild-de Sitter Spacetime: Properties of the Hadamard function'' 
\textit{Phys. Rev. D}
\textbf{111}
L021702 (2025).


\bibitem{G-A-E}
     M. Good, P. Anderson, C. Evans, ``Mirror reflections of a black hole'' \textit{Phys. Rev. D} \textbf{94} 065010 (2016).

  \bibitem{Bunch-Christensen-Fulling}
     T. Bunch, S. M. Christensen, S. A. Fulling,  ``Massive quantum field theory in two-dimensional Robertson Walker space-time'',
     \textit{Phys. Rev. D} \textbf{18} 4435 (1978).

 \bibitem{Christensen-Fulling}
    S. M. Christensen and S. A. Fulling, ``Trace anomalies and the Hawking effect'', \textit{Phys. Rev. D} \textbf{15} 2088 (1977).

     \bibitem{Christensen1976}
    S. Christensen.  `` Vacuum expectation value of the stress tensor in an arbitrary curved background: The covariant point-separation method", \textit{Physical Review D} \textbf{14} 2490 (1976).    
    \end{thebibliography}
\end{document}